\documentclass[useAMS,usenatbib]{mn2e}

\usepackage{graphicx,amssymb,verbatim,times}

\newcommand\lya{Ly$\alpha$}

\newcommand\ip{$i_{775}$}

\newcommand\zp{$z_{850}$}

\newcommand\vp{$V_{606}$}

\newcommand\h{$h^{-1}$}
\newcommand\idrops{$i$-dropouts}
\newcommand\apj{ApJ}
\newcommand\aj{AJ}
\newcommand\apjl{ApJL}
\newcommand\apjs{ApJS}
\newcommand\aap{A\&A}
\newcommand\mnras{MNRAS}
\newcommand\pasp{PASP}
\newcommand\pasj{PASJ}
\newcommand\nat{Nature}

\title[Galaxies, protoclusters and quasars at $z\sim6$]{$\Lambda$CDM
  predictions for galaxy protoclusters I: The relation between
  galaxies, protoclusters and quasars at $z\sim6$}
\author[R. A. Overzier et al.]{Roderik. A. Overzier$^{1}$\thanks{E-mail:overzier@mpa-garching.mpg.de (RAO)}, Qi Guo$^{1}$, Guinevere Kauffmann$^{1}$, Gabriella De Lucia$^{1}$,\and Rychard Bouwens$^{2}$, Gerard Lemson$^{3,4}$\\
  $^{1}$Max-Planck-Instit\"ut f\"ur Astrophysik, Karl-Schwarzschild-Str. 1, D-85748, Garching, Germany\\
  $^{2}$Astronomy Department, University of California, Santa Cruz, CA 95064, USA\\
  $^{3}$Astronomisches Rechen-Institut, Zentrum f\"ur Astronomie der Universit\"at Heidelberg, Moenchhofstr. 12-14, 69120 Heidelberg, Germany\\
  $^{4}$Max-Planck Institut f\"ur extraterrestrische Physik, Giessenbach Str., 85748 Garching, Germany}

\begin{document}

\date{}

\pagerange{\pageref{firstpage}--\pageref{lastpage}} \pubyear{2008}

\maketitle

\label{firstpage}

\begin{abstract}
  Motivated by recent observational studies of the environment of
  $z\sim6$ QSOs, we have used the Millennium Run (MR) simulations to
  construct a very large ($\sim4\degr\times4\degr$) mock redshift
  survey of star-forming galaxies at $z\sim6$. We use this simulated
  survey to study the relation between density enhancements in the
  distribution of \ip-dropouts and \lya\ emitters, and their relation
  to the most massive halos and protocluster regions at $z\sim6$. Our
  simulation predicts significant variations in surface density across
  the sky with some voids and filaments extending over scales of
  1$\degr$, much larger than probed by current surveys.  Approximately
  one third of all $z\sim6$ halos hosting $i$-dropouts brighter than
  $z$=26.5 mag ($\approx M^*_{UV,z=6}$) become part of $z=0$ galaxy
  clusters.  $i$-dropouts associated with protocluster regions are
  found in regions where the surface density is enhanced on scales
  ranging from a few to several tens of arcminutes on the sky. We
  analyze two structures of $i$-dropouts and \lya\ emitters observed
  with the Subaru Telescope and show that these structures must be the
  seeds of massive clusters-in-formation. In striking contrast, six
  $z\sim6$ QSO fields observed with HST show no significant
  enhancements in their \ip-dropout number counts. With the present
  data, we cannot rule out the QSOs being hosted by the most massive
  halos. However, neither can we confirm this widely used assumption.
  We conclude by giving detailed recommendations for the
  interpretation and planning of observations by current and future
  ground- and space based instruments that will shed new light on 
  questions related to the large-scale structure at $z\sim6$.
\end{abstract}


\section{Introduction}

\label{sec:intro}

\noindent
During the first decade of the third Millennium we have begun to put
observational constraints on the status quo of galaxy formation at
roughly one billion years after the Big Bang
\citep[e.g.][]{stanway03,yan04a,bouwens03,bouwens04a,bouwens06,dickinson04,malhotra05,shimasaku05,ouchi05,overzier06}.
Statistical samples of star-forming galaxies at $z=6$ -- either
selected on the basis of their large ({\it i}--{\it z}) color due to
the Lyman break redshifted to $z\sim6$ ({\it i-dropouts}), or on the
basis of the large equivalent width of Ly$\alpha$ emission
(\lya\ emitters) -- suggest that they are analogous to the population
of Lyman break galaxies (LBGs) found at $z\sim3-5$
\citep[e.g.][]{bouwens07}. A small subset of the \ip-dropouts
has been found to be surprisingly massive or old
\citep{dow05,yan06,eyles07}. The slope of the UV luminosity function
at $z=6$ is very steep and implies that low luminosity objects
contributed significantly to reionizing the Universe
\citep{yan04b,bouwens07,khochfar07,overzier08a}. Cosmological
hydrodynamic simulations are being used to reproduce the abundances as
well as the spectral energy distributions of $z=6$ galaxies. Exactly
how these objects are connected to local galaxies
remains a highly active area of research
\citep[e.g.][]{dave06,harford06,nagamine06,nagamine08,night06,finlator07,robertson07}.

The discovery of highly luminous quasi-stellar objects (QSOs) at
$z\sim6$ \citep[e.g.][]{fan01,fan03,fan04,fan06a,goto06,venemans07a}
is of equal importance in our understanding of the formation of the first massive black
holes and galaxies. \citet{gunn65} absorption troughs in their spectra
demarcate the end of the epoch of reionization
\citep[e.g.][]{fan01,white03,walter04,fan06b}.  Assuming that high
redshift QSOs are radiating near the Eddington limit, they contain
supermassive black holes (SMBHs) of mass $\sim10^9$ $M_\odot$
\citep[e.g.][]{willott03,barth03,vestergaard04,jiang07,kurk07}.  
The spectral properties of most $z\sim6$ QSOs in the rest-frame UV, optical, IR and
X-ray are similar to those at low redshift, suggesting that massive,
and highly chemically enriched galaxies were vigorously forming stars
and SMBHs less than one billion years after the Big Bang
\citep[e.g.][]{bertoldi03,maiolino05,jiang06,wang07}.

Hierarchical formation models and simulations can reproduce the
existence of such massive objects at early times
\citep[e.g.][]{haiman01,springel05a,begelman06,volonteri06,li07,narayan07},
provided however that they are situated in extremely massive halos.  
Large-scale gravitational clustering is a powerful method for
estimating halo masses of quasars at low redshifts, but cannot 
be applied to $z\sim6$ QSOs because there are too few systems known. Their
extremely low space density determined from the Sloan Digital Sky
Survey (SDSS) of $\sim$1 Gpc$^{-3}$ (comoving) implies a (maximum)
halo mass of $\cal{M}$$_{halo}\sim10^{13}$ $M_\odot$ \citep{fan01,li07}. A
similar halo mass is obtained when extrapolating from the ($z=0$)
relationship between black hole mass and bulge mass of
\citet{magorrian98}, and using $\Omega_M/\Omega_{bar}\gtrsim10$
\citep{fan01}.  
Because the descendants of the most massive
halos at $z\sim6$ may evolve into halos of $>10^{15}$ $M_\odot$ at
$z=0$ in a $\Lambda$CDM model, 
\citep[e.g.][but see \citet{delucia07}, \citet{trenti08} and
  Sect. \ref{sec:qso} of this paper]{springel05a,suwa06,li07}, it is
believed that the QSOs trace highly biased regions that may give birth to the most
massive present-day galaxy clusters. If this is true, the small-scale
environment of $z\sim6$ QSOs may be expected to show a significant
enhancement in the number of small, faint galaxies. These galaxies may
either merge with the QSO host galaxy, or may form the first stars and
black holes of other (proto-)cluster galaxies.  

Observations carried out with the Advanced Camera for Surveys (ACS) on
the {\it Hubble Space Telescope} (HST), allowed a rough measurement of
the two-dimensional overdensities of faint \ip-dopouts detected towards the QSOs
J0836+0054 at $z=5.8$ \citep{zheng06} and J1030+0524 at $z=6.28$
\citep{stiavelli05}.  Recently \citet{kim08} presented results from a
sample of 5 QSO fields, finding some to be overdense and some to be
underdense with respect to the HST/ACS Great Observatories Origins
Deep Survey (GOODS).
\citet{priddey07} find enhancements in the number counts of sub-mm
galaxies. Substantial overdensities of \idrops\ and \lya\ emitters
have also been found in non-QSO fields
\citep[e.g.][]{shimasaku03,ouchi05,ota08}, suggesting that massive
structures do not always harbour a QSO, which may be explained by
invoking a QSO duty-cycle. At $z\sim2-5$, significant excesses of
star-forming galaxies have been found near QSOs
\citep[e.g.][]{djorgovski03,kashikawa07}, radio galaxies
\citep[e.g.][]{miley04,venemans07b,overzier08b}, and in random fields
\citep{steidel98,steidel05}.  Although the physical interpretation of
the measurements is uncertain, these structures are believed to be
associated with the formation of clusters of galaxies.

The idea of verifying the presence of massive structures
at high redshift through the clustering of small galaxies around them
has recently been explored by, e.g., \citet{munoz08a} using the
excursion set formalism of halo growth \citep{zentner07}.  However,
the direct comparison between models or simulations and observations
remains difficult, mainly because of complicated observational
selection effects. This is especially true at high redshift. 
In order to investigate how a wide variety of galaxy overdensities 
found in surveys at $z\simeq2-6$ are related to cluster formation, we
have carried out an analysis of the progenitors of galaxy clusters in
a set of cosmological $N$-body simulations. Our results will be
presented in a series of papers. In Paper I, we use the Milennium Run Simulations
\citep{springel05a} to simulate a large mock survey of galaxies at
$z\sim6$ to derive predictions for the properties of the progenitors
of massive galaxy clusters, paying particular attention to the details
of observational selection effects.  We will try to answer the following
questions:\\

(i) Where do we find the present-day descendants of the {\it i}-dropouts?

(ii) What are the typical structures traced by {\it i}-dropouts and
\lya\ emitters in current surveys, and how do they relate to protoclusters?

(iii) How do we unify the (lack of excess) number counts observed in
QSO fields with the notion that QSOs are hosted by the most massive halos at $z\sim6$?\\

The structure of the present paper is as follows. We describe the
simulations, and construction of our mock {\it i-dropout} survey in
Section 2. Using these simulations, we proceed to address the main
questions outlined above in Sections 3--5. We conclude the paper
with a discussion (Section 6), an overview of recommendations for future
observations (Section 7), and a short summary (Section 8) of the main results.

\section{Simulations}

\subsection{Simulation description}

\noindent
We use the semi-analytic galaxy catalogues that are based on the
Millennium Run (MR) dark matter simulation of \citet{springel05a}.
Detailed descriptions of the simulations and the semi-analytic
modeling have been covered extensively elsewhere, and we kindly refer
the reader to those works for more information \citep[e.g.][and references
 therein]{kauffmann99,springel05a,croton06,lemson06a,delucia04,delucia07}.

The dark matter simulation was performed with the
cosmological simulation code GADGET--2 \citep{springel05b}, and
consisted of $2160^3$ particles of mass $8.6\times10^8$ $h^{-1}$
$M_\odot$ in a periodic box of 500 $h^{-1}$ Mpc on a side.  The
simulations followed the gravitational growth as traced by these
particles from $z=127$ to $z=0$ in a $\Lambda$CDM
cosmology ($\Omega_m=0.25$, $\Omega_\Lambda=0.75$, $h=0.73$, $n=1$,
$\sigma_8=0.9$) consistent with the WMAP year 1 data
\citep{spergel03}. The results were stored at 64 
epochs (``snapshots''), which were used to construct a detailed halo
merger tree during postprocessing, by identifying all resolved dark
matter halos and subhalos, and linking all progenitors and descendants
of each halo.
\begin{figure}
\begin{center}
\includegraphics[width=\columnwidth]{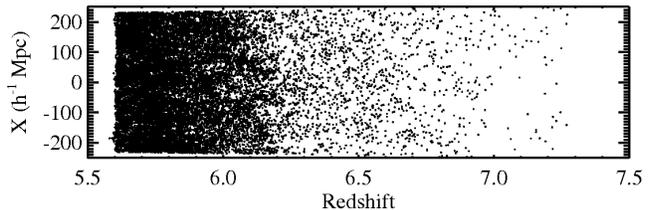}
\end{center}
\caption{\label{fig:zx} Redshift versus the (co-moving) $X$-coordinate
  for all objects within a slice of width $\Delta Y=250$ $h^{-1}$
  Mpc along the $Y$-axis.}
\end{figure}

Galaxies were modeled by applying semi-analytic prescriptions of
galaxy formation to the stored halo merger trees. The techniques and
recipes include gas cooling, star formation, reionizaton heating,
supernova feedback, black hole growth, and ``radio-mode'' feedback
from galaxies with a static hot gas atmosphere, and are described in
\citet{croton06}. The photometric properties of galaxies are then
modeled using stellar population synthesis models, including a simple
dust model.  Here we use the updated models `delucia2006a' of
\citet{delucia07} that have been made publicly available through an
advanced database structure on the MR
website\footnote{http://www.mpa-garching.mpg.de/millennium/} \citep{lemson06b}.

\begin{figure}
\begin{center}
\includegraphics[width=\columnwidth]{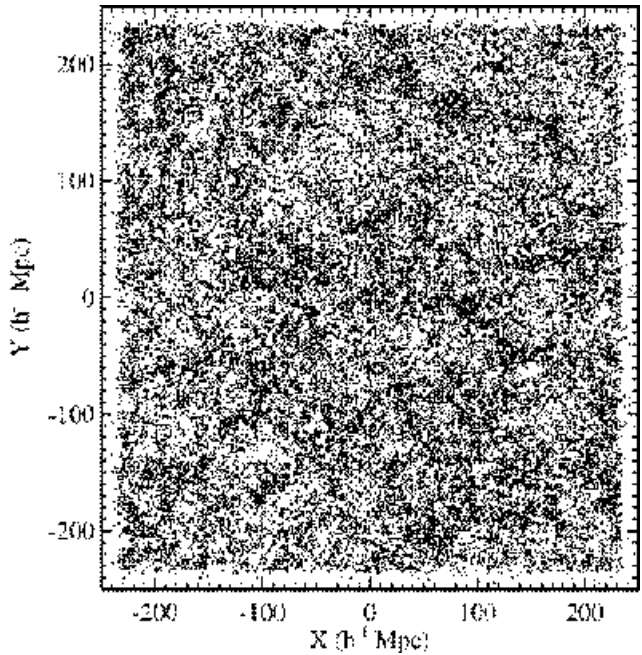}
\end{center}
\caption{\label{fig:xy} The simulation box showing the positions in
  co-moving coordinates of all objects identified as \ip-dropout
  galaxies to \zp$=$27.0 mag.}
\end{figure}

\subsection{Construction of a large mock survey at $z\sim6$}
\label{sec:makemock}

\noindent
We used the discrete MR snapshots to create a large, continous mock
survey of \ip-dropout galaxies at $z\sim6$.  The general principle of
transforming a series of discrete snapshots into a
mock pencil beam survey entails placing a virtual observer
somewhere in the simulation box at $z=0$ and carving out all galaxies
as they would be observed in a pencil beam survey along that
observer's line of sight.  This technique has been described in great
detail in \citet{blaizot05} and \citet{kitzbichler07}. In general, one 
starts with the snapshot $i=63$ at $z=0$ and records the positions,
velocities and physical properties of all galaxies within the cone out
to a comoving distance corresponding to that of the next snapshot. For
the next segment of the cone, one then use the properties as recorded in
snapshot $i=62$, and so on. The procedure relies on the reasonable assumption
that the large-scale structure (positions and velocities of galaxies)
evolves relatively slowly between snapshots. By replicating the simulation
box along the lightcone and limiting the opening angle of the cone,
one can in principle construct unique lightcones out to very high
redshift without crossing any region in the simulation box more than
once.  The method is 
straightforward when done in comoving coordinates in a flat cosmology
using simple Euclidean geometry \citep{kitzbichler07}.

Because the comoving distances or redshifts of galaxies recorded at a
particular snapshot do not correspond exactly to their effective
position along the lightcone, we need to correct their magnitudes by
interpolating over redshift as follows:
\begin{equation}
M_{cor}[z(d)]=M(z_i)+\frac{dM}{dz}[z(d)-z_i)],
\end{equation}
where $M_{cor}[z(d)]$ is the observer-frame absolute magnitude at the
observed redshift, $z(d)$ (including peculiar velocities along the
line of sight), $M(z_i)$ is the magnitude at redshift $z_i$
corresponding to the $i$th snapshot, and $dM/dz$ is
the first order derivative of the observer-frame absolute
magnitude. The latter quantity is calculated for each galaxy by
placing it at neighbouring snapshots, and ensures that the
$K$-correction is taken into account \citep{blaizot05}. Finally, we apply
the mean attenuation of the intergalactic medium using \citet{madau95}
and calculate the observer-frame apparent magnitudes in each filter.

In this paper, we use the fact that the selection of $z\sim6$
galaxies through the {\it i-dropout} technique is largely free of contamination
from objects at lower (and higher) redshift \citep[][]{bouwens06}
provided that the observations are deep enough. 
Because the transverse size of the MR simulation box (500 \h\ Mpc)
corresponds to a comoving volume between $z\approx5.6$ and
$z\approx7.3$ (the typical redshift range of \idrops\ surveys) we can
use the three simulation
snapshots centered at $z=5.7$, $z=6.2$ and $z=6.7$ to create a 
mock survey spanning this volume, while safely
neglecting objects at other redshifts.
\begin{figure*}
\includegraphics[width=0.45\textwidth]{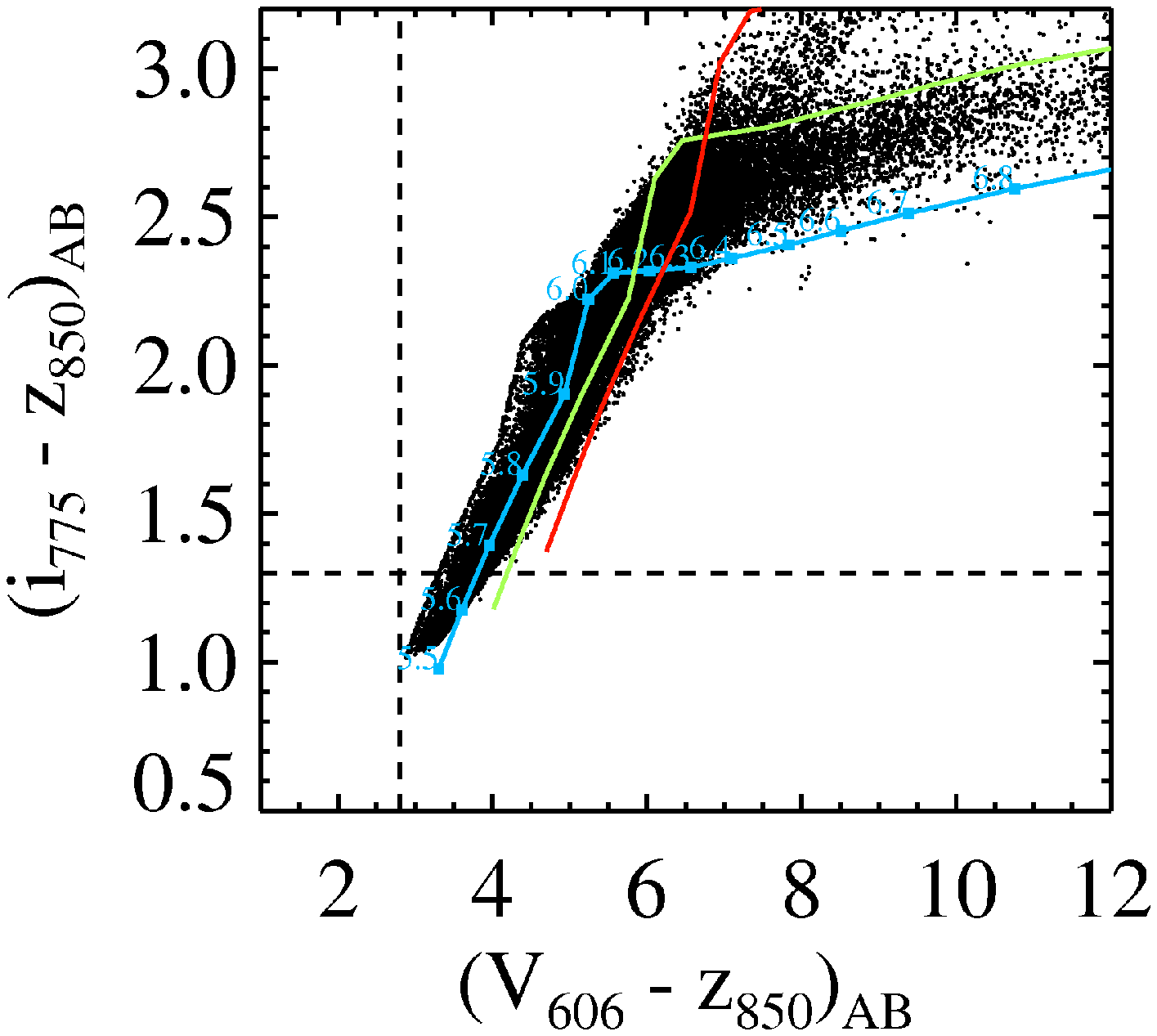}
\hspace{1cm}
\includegraphics[width=0.43\textwidth]{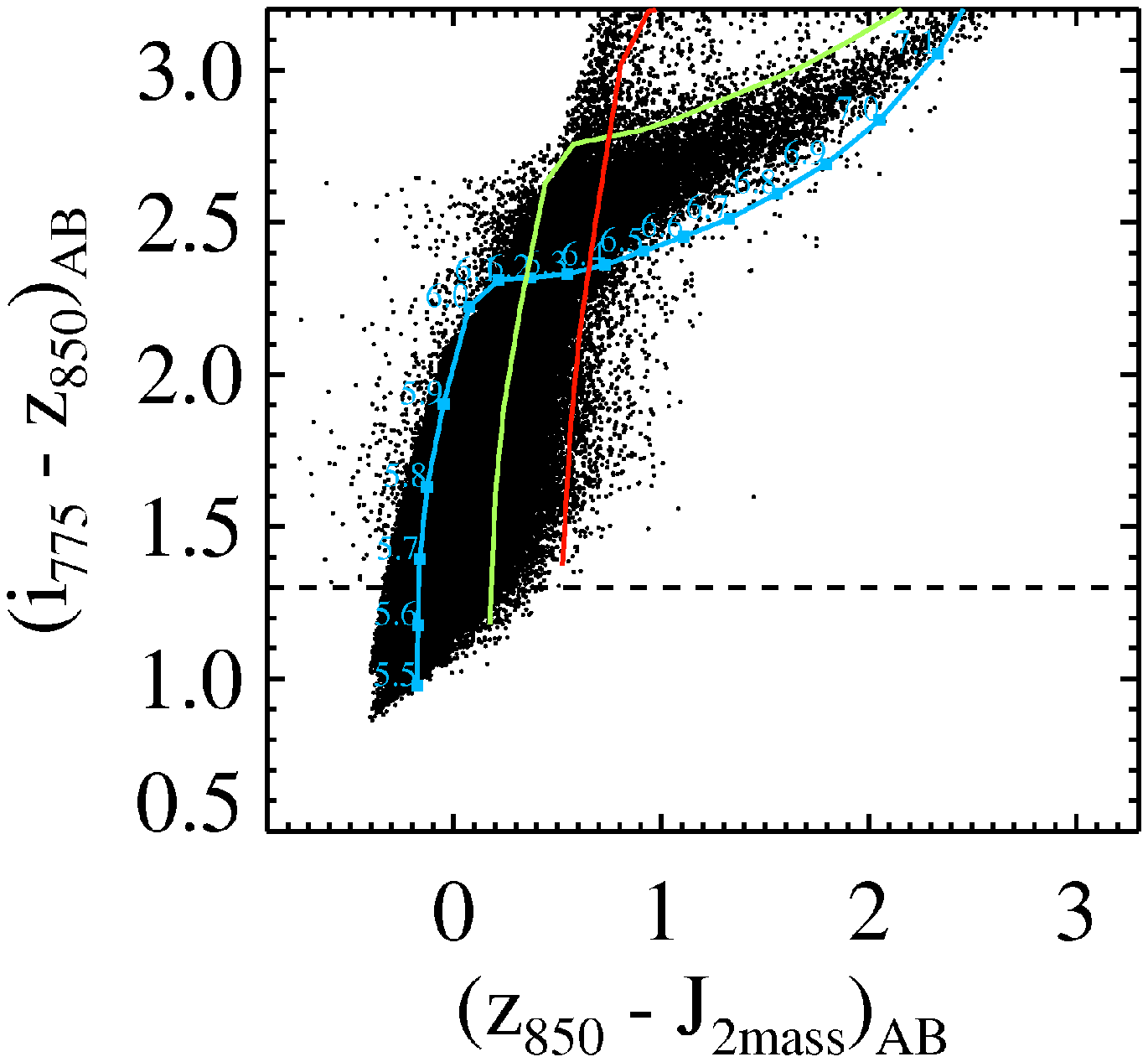}
\caption{\label{fig:color}Colour-colour diagrams of the MR mock
  \ip-dropout survey. To guide the eye we have indicated tracks
  showing the colours of a 100 Myr old continuous starburst model from
  \citet{bc03} for different amounts of reddening in $E(B-V)$ of 0.0
  (blue), 0.2 (green), and 0.4 (red). Redshifts are indicated along the
  zero-reddening track. Only objects at $z>5.6$ are included in the
  simulations, as \ip-dropouts surveys have been demonstrated to have
  very little contamination (see text for details).}
\end{figure*}

We extracted galaxies from the MR database by selecting the 
$Z-$axis of the simulation box to lie along the line-of-sight of our mock
field. In order to compare with the deepest
current surveys, we calculated the apparent magnitudes in the HST/ACS
\vp, \ip\ and \zp\ filters and the 2MASS $J$, $H$ and $K_S$
filters. We derived observed redshifts from the comoving distance
along the line of sight (including the effect of peculiar velocities),
applied the $K$-corrections and IGM absorption, and calculated the
apparent magnitudes in each band. Fig. \ref{fig:zx} shows the spatial
$X$-coordinate versus the redshift of objects in the simulated
lightcone.  Fig. \ref{fig:xy} shows the entire simulated volume
projected along the $Z$- or redshift axis. These figures show that
there exists significant filamentary and strongly clustered
substructure at $z\approx6$, both parallel and perpendicular to the
line-of-sight.

Our final mock survey has a comoving volume of $\sim0.3$ Gpc$^3$, and
spans an area of $4.4\degr\times4.4\degr$ when projected onto the
sky. It contains $\sim1.6\times10^5$ galaxies at $z=5.6-7.3$ with 
$z\le27.5$ mag (corresponding to an absolute magnitude\footnote{The
  rest-frame absolute magnitude at 1350\AA\ is defined as
  $M_{1350\AA}\simeq m_z - 5\log_{10}(d_L/\mathrm{10pc}) + 2.5\log_{10}(1+z)$} 
of $M_{UV,AB}\simeq-19.2$ mag, about one mag
below $M^*_{UV,z=6}$).  For comparison and future reference, we list
the main $i$-dropout surveys together with their areal coverage and
detection limit in Table \ref{tab:surveys}.

\subsection{Colour-colour selection}

\noindent
In the left panel of Fig. \ref{fig:color} we show the
$V_{606}-z_{850}$ vs. $i_{775}-z_{850}$ colour-colour diagram for all
objects satisfying $z\le27.0$.  The $i$-dropouts populate a region in
colour-colour space that is effectively isolated from lower redshift
objects using a simple colour cut of
$i_{775}-z_{850}\gtrsim1.3-1.5$. Note that although our simulated
survey only contains objects at $z>5.6$, it has been shown
\citep{stanway03,dickinson04,bouwens04a,bouwens06} that this colour
cut is an efficient selection criterion for isolating starburst
galaxies at $z\sim6$ with blue \zp$-J$ colours (see right panel of
Fig. \ref{fig:color}). For reference, we have overplotted colour
tracks for a 100 Myr old, continuous star formation model as a
function of redshift; different colour curves show results for
different amounts of reddening by
dust. As can be seen, these simple models span the region of
colour-colour space occupied by the MR galaxies. At $z<6$, galaxies
occupy a tight sequence in the plane. At $z>6$, objects fan out
because the  $V_{606}-z_{850}$ colour changes strongly as a function
of redshift, while the $i_{775}-z_{850}$ colour is more sensitive to
both age and dust reddening. 
Because of the possibility of intrinsically red interlopers at $z\sim1-3$, the
additional requirement of a non-detection in $V_{606}$, or a very red
$V_{606}-z_{850}\gtrsim3$ colour, if available, is often included in
the selection\footnote{The current paper uses magnitudes and colours
  defined in the HST/ACS \vp\ip\zp\ filter system in order to compare
  with the deepest surveys available in literature. Other works based
  on groundbased data commonly use the SDSS-based $r^\prime i^\prime
  z^\prime$ filterset, but the differences in colours are minimal.}.
Because the selection based on $i_{775}-z_{850}\gtrsim1.3$ introduces
a small bias against objects having strong \lya\ emission at
$z\lesssim6$ \citep{malhotra05,stanway07}, we have statistically
included the effect of \lya\ on our sample selection by randomly
assigning \lya\ with a rest-frame equivalent width of 30\AA\ to 25\%
of the galaxies in our volume, and recalculating the $i_{775}-z_{850}$
colours. The inclusion of \lya\ leads to a reduction in the number of
objects selected of $\sim3$\% \citep[see also][]{bouwens06}.

\subsection{$i$-dropout number densities}

\noindent
In Table \ref{tab:surfdens} we list the surface densities of
\ip-dropouts selected in the MR mock survey as a function of limiting
\zp-magnitude and field size.
For comparison, we calculated the surface densities for regions having
areas comparable to some of the main \ip-dropout surveys: the SDF (876
arcmin$^2$), two GOODS fields (320 arcmin$^2$), a single GOODS field
(160 arcmin$^2$), and a HUDF-sized field (11.2 arcmin$^2$). The errors
in Table \ref{tab:surfdens} indicate the $\pm1\sigma$ deviation
measured among a large number of similarly sized fields selected from
the mock survey, and can be taken as an estimate of the influence of
(projected) large-scale structure on the number counts (usually
referred to as ``cosmic variance'' or ``field-to-field variations'').
At faint magnitudes, the strongest observational constraints on the
\ip-dropout density come from the HST surveys. Our values for a GOODS-sized
survey are 105, 55 and 82\% of the values given by the most recent
estimates by B07 for limiting \zp\ magnitudes of 27.5, 26.5 and 26.0 mag, respectively, and
consistent within the expected cosmic variance allowed by our mock
survey.  Because the total area surveyed by B06 is about
200$\times$\ smaller than our mock survey, we also compare our results
to the much larger SDF from \citet{ota08}. At $z=26.5$ mag the
number densities of $\sim$0.18 arcmin$^{-2}$ derived from both the
real and mock surveys are perfectly consistent.

Last, we note that in order to achieve agreement between the observed
and simulated number counts at $z\sim6$, we did not require any tweaks
to either the cosmology \citep[e.g., see][for the effect of different
WMAP cosmologies]{wang08}, or the dust model used \citep[see][for
alternative dust models better tuned to high redshift
galaxies]{kitzbichler07,guo08}. This issue may be further
investigated in a future paper.

\subsection{Redshift distribution}

\begin{figure}
\begin{center}
\includegraphics[width=\columnwidth]{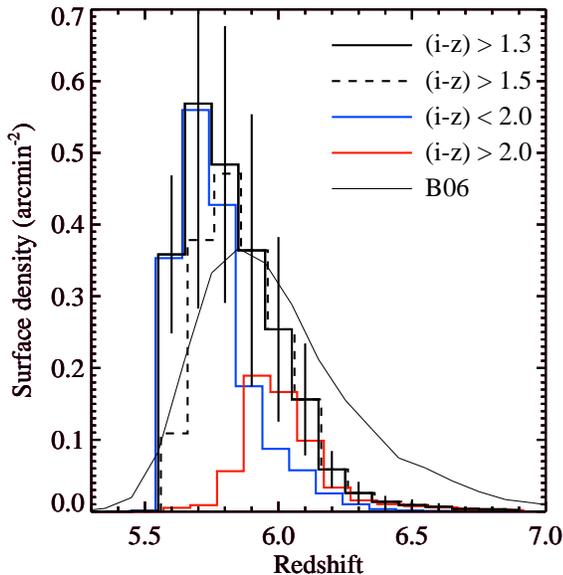}
\end{center}
\caption{\label{fig:nz2}Redshift histograms derived from the MR mock
  $i$-dropout survey at the depth of \zp=27.5 mag using the selection
  criteria \ip--\zp$>1.3$ (thick solid line, error bars indicate the
  1$\sigma$ scatter expected among GOODS-sized fields), \ip--\zp$>1.5$
  (dashed line), \ip--\zp$<2.0$ (blue line), and \ip--\zp$>2.0$ (red
  line).  The thin solid line indicates the model redshift
  distribution from B06 based on the HUDF.}
\end{figure}

\noindent
In Fig. \ref{fig:nz2} we show the redshift distribution of the full
mock survey (thick solid line), along with various subsamples selected
according to different \ip--\zp\ colour cuts that we will refer to
later on in this paper. The standard selection of \ip--\zp$>$1.3
results in a distribution that peaks at $z\approx5.8$. We have also
indicated the expected scatter resulting from cosmic variance on the
scale of GOODS-sized fields (error bars are 1$\sigma$).  Some, mostly
groundbased, studies make use of a more stringent cut of
\ip--\zp$>1.5$ to reduce the chance of foreground interlopers (dashed
histogram). Other works have used colour cuts of \ip--\zp$\lesssim2$
(blue histogram) and \ip--\zp$\gtrsim2$ (red histogram) in order to
try to extract subsamples at $z\lesssim6$ and $z\gtrsim6$,
respectively. As can be seen in Fig. \ref{fig:nz2}, such cuts are
indeed successful at broadly separating sources from the two redshift
ranges, although the separation is not perfectly clean due to the
mixed effects of age, dust and redshift on the \ip--\zp\ colour.  For
reference, we have also indicated the model redshift distribution from
B06 (thin solid line). This redshift distribution was derived for a
much fainter sample of \zp$\lesssim$29 mag, which explains in part the
discrepancy in the counts at $z\gtrsim6.2$. Evolution across the
redshift range will furthermore skew the actual redshift distribution
toward lower values \citep[see discussion in][]{munoz08b}. This is not
included in the B06 model, and its effect is only marginally taken
into account in the MR mock survey due to the relatively sparse
snapshot sampling across the redshift range. Unfortunately, the exact
shape of the redshift distribution is currently not very well
constrained by spectroscopic samples \citep{malhotra05}. A more
detailed analysis is beyond the scope of this paper, and we conclude
by noting that the results presented below are largely independent of
the exact shape of the distribution.
 
\subsection{Physical properties of $i$-dropouts}

\begin{figure}
\begin{center}
\includegraphics[width=\columnwidth]{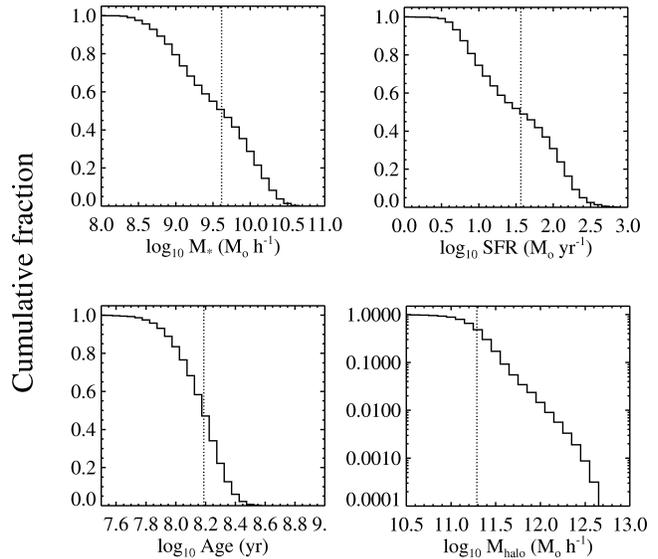}
\end{center}
\caption{\label{fig:z6props}Physical properties of $i$-dropouts in the
  MR mock survey satisfying $z\le26.5$ mag.  We plot the cumulative
  fractions of galaxies with stellar masses, star formation rates,
  stellar ages and halo masses greater than a given value. {\it Top left:}
  Distribution of stellar masses. The median stellar mass is
  $\sim4\times10^{9}$ $M_\odot$ h$^{-1}$ (dotted line). {\it Top
    right:} Distribution of SFRs. The median SFR is $\sim30$ $M_\odot$
  yr$^{-1}$ (dotted line).  {\it Bottom left:} Distribution of
  mass-weighted ages. The median age is $\sim$160 Myr (dotted
  line). {\it Bottom right:} Distribution of halo masses. The median
  halo mass is $\sim2\times10^{11}$ $M_\odot$ h$^{-1}$ (dotted line).}
\end{figure}

\noindent
Although a detailed study of the successes and failures in the
semi-analytic modeling of galaxies at $z\sim6$ is not the purpose of
our investigation, we believe it will be instructive for the reader if
we at least summarize the main physical properties of the model
galaxies in our mock survey.
Unless stated otherwise, throughout this paper we will limit our
investigations to $i$-dropout samples having a limiting magnitude of
\zp=26.5 mag\footnote{For reference: a \zp\ magnitude of $\simeq$26.5
  mag for an unattenuated galaxy at $z\simeq6$ would correspond to a SFR
  of $\simeq$7 $M_\odot$ yr$^{-1}$, under the widely used assumption of a
  0.1--125$M_\odot$ Salpeter initial mass function 
and the conversion factor between SFR and the
  rest-frame 1500\AA\ UV luminosity of $8.0\times10^{27}$ erg s$^{-1}$
  Hz$^{-1}$ / $M_\odot$ yr$^{-1}$ as given by \citet{madau98}.}, comparable to $M^*_{UV}$ at $z=6$
\citep[see][]{bouwens07}.  This magnitude typically corresponds to
model galaxies situated in dark matter halos of at least 100 dark
matter particles ($\sim10^{11}$ M$_\odot$ h$^{-1}$). This ensures that
the evolution of those halos and their galaxies has been traced for
some time prior to the snapshot from which
the galaxy was selected. In this way, we ensure that the physical quantities derived from
the semi-analytic model are relatively stable against
snapshot-to-snapshot fluctuations.  A magnitude limit of \zp=26.5 mag
also conveniently corresponds to the typical depth that can be
achieved in deep groundbased surveys or relatively shallow HST-based
surveys.

\begin{figure}
\begin{center}
\includegraphics[width=\columnwidth]{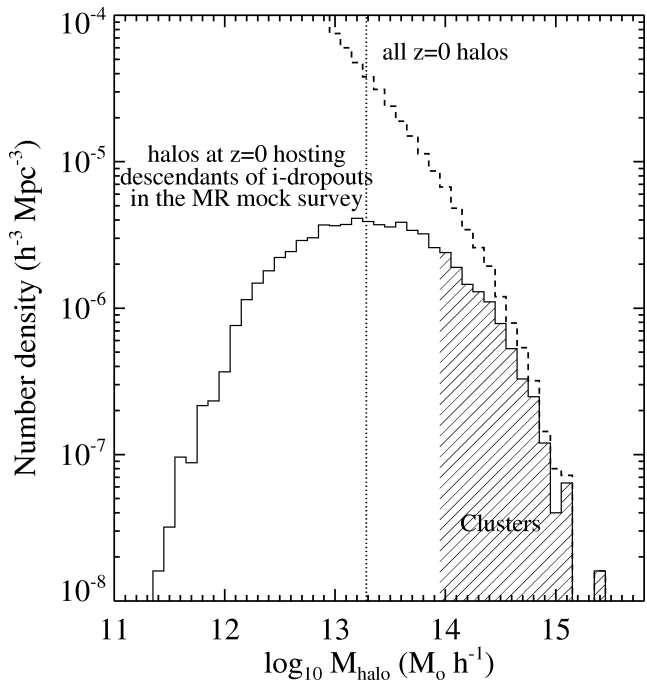}
\end{center}
\caption{\label{fig:z0halos}Number density versus halo mass of the
  $z=0$ dark matter halos hosting descendants of $i$-dropouts at
  $z\sim6$ to a limiting depth of \zp$\le$26.5 mag. The median
  $i$-dropout descendant halo mass is a few times $10^{13}$ $M_\odot$
  (dotted line). The halo mass function of all MR halos at $z=0$ is
  indicated by the dashed line. The mass range occupied by the halos
  associated with galaxy clusters is indicated by the hatched region.}
\end{figure}

In Fig. \ref{fig:z6props} we plot the cumulative distributions of the
stellar mass (top left), SFR (top right), and stellar age (bottom
left) of the \ip-dropouts in the mock survey.  The median stellar mass is
$\sim5\times10^{9}$ $M_\odot$ h$^{-1}$, and about 30\% of galaxies
have a stellar mass greater than $10^{10}$ $M_\odot$. The median SFR
and age are $\sim30$ $M_\odot$ yr$^{-1}$ and $\sim$160 Myr,
respectively, with extrema of $\sim500$ $M_\odot$ yr$^{-1}$ and
$\sim$400 Myr. These results are in general agreement with several
studies based on modeling the stellar populations of limited samples
of \ip-dropouts and \lya\ emitters for which deep observations
with {\it HST} and {\it Spitzer} exist. \citet{yan06} have analyzed 
a statistically robust sample and find stellar masses
ranging from $\sim1\times10^8$ $M_\odot$ for IRAC-undetected sources
to $\sim7\times10^{10}$ $M_\odot$ for the brightest 3.6$\mu$m sources,
and ages ranging from $<$40 to 400 Myr \citep[see also][for
  additional comparison data]{dow05,eyles07,lai07}.  We also point out
that the maximum stellar mass of $\sim7\times10^{10}$ $M_\odot$ found
in our mock survey (see top left panel) is comparable to the most
massive $i$-dropouts found, and that ``supermassive'' galaxies having
masses in excess of $\gtrsim10^{11}$ $M_\odot$ are absent in both the
simulations and observations \citep{mclure06}. Last, in the bottom
right panel we show the distribution of the masses of the halos
hosting the model $i$-dropouts. The median halo mass is
$\sim3\times10^{11}$ $M_\odot$. Our results are in the range of
values reported by \citet{overzier06} and \citet{mclure08} based on
the angular correlation function of large \ip-dropout samples, but we
note that halo masses are currently not very well constrained by
the observations.

\begin{figure}
\begin{center}
\includegraphics[width=\columnwidth]{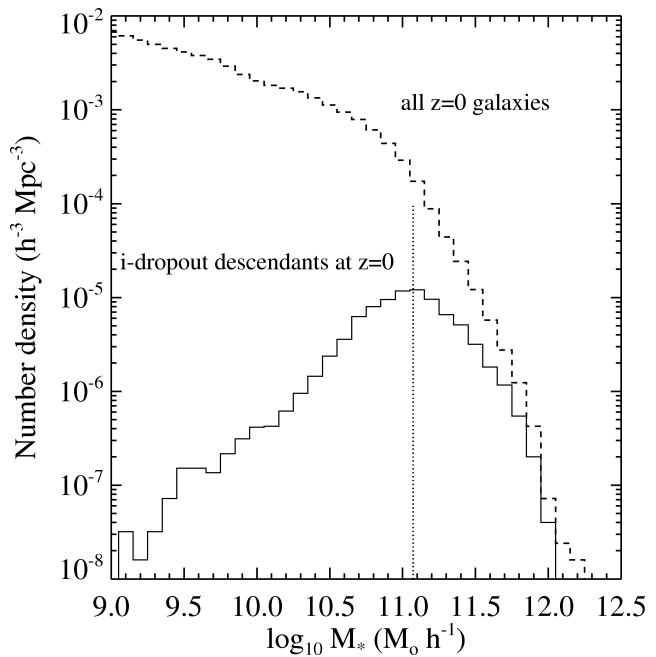}
\end{center}
\caption{\label{fig:z0galaxies}Number density versus stellar mass of
  the galaxies at $z=0$ that have at least one $i$-dropout progenitor
  at $z\sim6$. The median descendant mass is $\sim10^{11}$ $M_\odot$
  (dotted line). The distribution of stellar mass of all MR $z=0$
  galaxies is indicated for comparison (dashed line).}
\end{figure}

\section{The relation between $i$-dropouts and (proto-)clusters}

\noindent
In this Section we study the relation between local overdensities in
the $i$-dropout distribution at $z\sim6$ and the sites of cluster
formation.  
Throughout this paper, a galaxy cluster is defined as being all
galaxies belonging to a bound dark matter halo having a dark matter
mass\footnote{The `tophat' mass, $\cal{M}$$_{tophat}$, is the mass
  within the radius at which 
  the halo has an overdensity corresponding to the value at
  virialisation in the top-hat collapse model \citep[see][]{white01}.} of
$\cal{M}$$_{tophat}\ge10^{14}$ h$^{-1}$ $M_\odot$ at $z=0$.  In the MR we find
2,832 unique halos, or galaxy clusters, fulfilling this condition, 21
of which can be considered supermassive ($\cal{M}$$_{tophat}\ge10^{15}$
h$^{-1}$ $M_\odot$). Furthermore, a proto-cluster galaxy is defined as
being a galaxy at $\sim6$ that will end up in a galaxy cluster at
$z=0$.  Note that these are trivial definitions given the database
structure of the Millennium Run simulations, in which galaxies and
halos at any given redshift can be related to their progenitors and descendants at
another redshift \citep{lemson06b}. 

\begin{figure*}
\begin{center}
\includegraphics[width=0.8\textwidth]{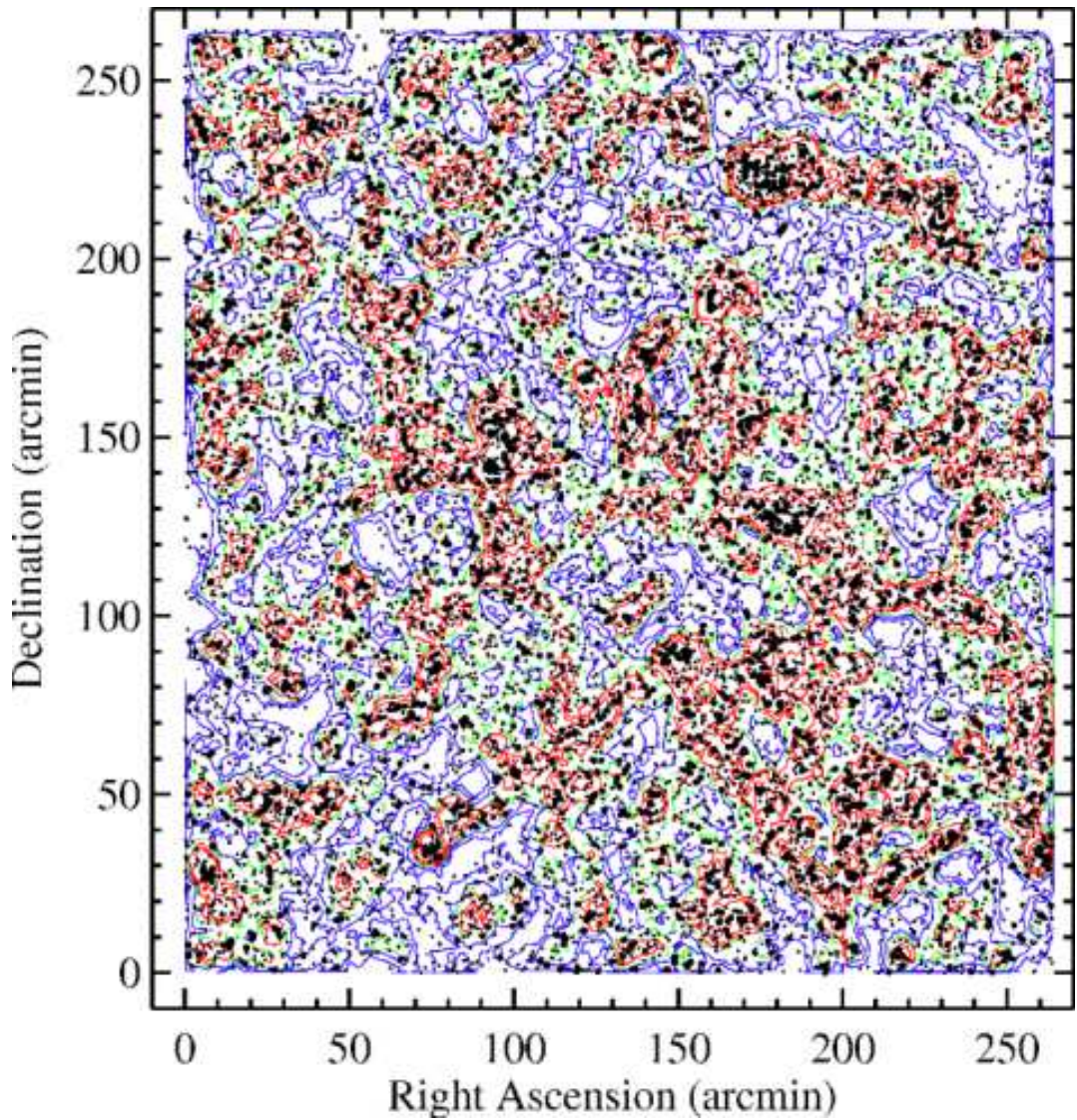}
\end{center}
\caption{\label{fig:field}Projected distribution on the sky of the
  $z\sim6$ $i$-dropouts selected from the MR mock survey according to
  the criteria \ip--\zp$>$1.3 and $z\le26.5$ mag (small and large
  points). Contours indicate regions of equal density, defined as
  $\delta\Sigma_{5\arcmin}\equiv(\Sigma_{5\arcmin}-\bar{\Sigma}_{5\arcmin})/\bar{\Sigma}_{5\arcmin}$,
  $\Sigma_{5\arcmin}$ and $\bar{\Sigma}_{5\arcmin}$ being the
  local and mean surface density measured in circular cells of
  5\arcmin\ radius. Over- and underdense regions of
  $\delta\Sigma_{5\arcmin}=\pm[0.25,0.5,1.0]$ are shown in red and
  blue contours, respectively. The mean density
  ($\delta\Sigma_{5\arcmin}=0$) is indicated by the green dashed
  contour. Large black points mark proto-cluster galaxies that end up in
  galaxy clusters at $z=0$.}
\end{figure*}

\subsection{Properties of the $z=0$ descendants of $i$-dropouts}

\noindent
In Fig. \ref{fig:z0halos} we plot the distribution of number densities of the central
halos that host the $z=0$ descendants of the $i$-dropouts in our mock
survey as a function of the halo mass. The median halo mass hosting
the $i$-dropout descendants at $z=0$ is $3\times10^{13}$ $M_\odot$
h$^{-1}$ (dotted line). For comparison we indicate the mass
distribution of all halos at $z=0$ (dashed line). The plot shows that
the fraction of halos that can be traced back to a halo hosting an
$i$-dropout at $z\sim6$ is a strong function of the halo mass at
$z=0$. 45\% of all cluster-sized halos at $z=0$ (indicated by the
hatched region) are related to the descendants of halos hosting
$i$-dropouts in our mock survey, and 77\% of all clusters at $z=0$
having a mass of $\cal{M}$$>7\times10^{14}$ $M_\odot$ h$^{-1}$ can be traced
back to at least one progenitor halo at $z\sim6$ hosting an
$i$-dropout. This implies that the first seeds of galaxy clusters are
already present at $z\sim6$. In addition, many $i$-dropout galaxies
and their halos may merge and end up in the same descendant structures
at $z=0$, which was not accounted for in our calculation above where we
only counted unique halos at $z=0$. In fact, about $\sim$34\%
($\sim$2\%) of all $i$-dropouts (\zp$\le$26.5) in the mock survey will
end up in clusters of mass $>1\times10^{14}$ ($>7\times10^{14}$)
$M_\odot$ h$^{-1}$ at $z=0$. This implies that roughly one third of
all galaxies one observes in a typical $i$-dropout survey can be
considered ``proto-cluster'' galaxies. The plot further shows that the
majority of halos hosting $i$-dropouts at $z\sim6$ will evolve into
halos that are more typical of the group environment. This is similar
to the situation found for Lyman break or dropout galaxies at lower
redshifts \citep{ouchi04}.

In Fig. \ref{fig:z0galaxies} we plot the stellar mass distribution of
those $z=0$ galaxies that host the descendants of the
$i$-dropouts.  The present-day descendants are found in galaxies
having a wide range of stellar masses ($\cal{M}$$_*\simeq10^{9-12}$ $M_\odot$),
but the distribution is skewed towards the most massive galaxies in
the MR simulations. The median stellar mass of the descendants 
is $\sim10^{11}$ $M_\odot$ (dotted line in Fig. \ref{fig:z0galaxies}).

\subsection{Detecting proto-clusters at $z\sim6$}
\label{sec:detclus}

\begin{figure}
\begin{center}
\includegraphics[width=\columnwidth]{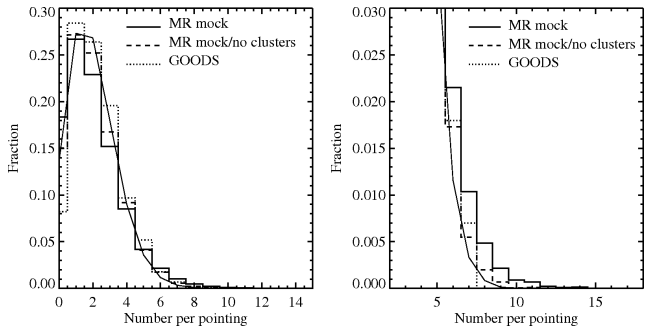}
\end{center}
\caption{\label{fig:npointing}Counts-in-cells frequency distribution
  of the $i$-dropouts shown in Fig. \ref{fig:field}, based on 20,000
  randomly placed ACS-sized fields of $3.4\arcmin\times3.4\arcmin$.
  The panel on the right shows a zoomed-in view to give a better sense
  of the small fraction of pointings having large numbers of
  $i$-dropouts. In both panels, the thick solid line indicates the
  frequency distribution of the full MR mock survey. The dashed line
  indicates how the distribution changes if we ``disrupt'' all
  protocluster regions of Fig. \ref{fig:field} by randomizing the
  positions of the galaxies marked as proto-cluster galaxy. The dotted
  line indicates the frequency distribution of a large sample of
  $i$-dropouts selected from the GOODS survey by B06 using identical
  selection criteria. Thin solid lines indicate the Poisson distribution
  for a mean of 2 $i$-dropouts per pointing.}
\end{figure}

\noindent
We will now focus on to what extent local overdensities in the
$i$-dropout distribution at $z\approx6$ may trace the progenitor seeds of
the richest clusters of galaxies in the present-day Universe. 
In Fig. \ref{fig:field} we plot the sky distribution of the
$i$-dropouts in our $4.4\degr\times4.4\degr$ MR mock survey (large and
small circles). Large circles indicate those $i$-dropouts 
identified as proto-cluster galaxies. We have plotted
contours of $i$-dropout surface density,
$\delta_{\Sigma,5^\prime}\equiv(\Sigma_{5^\prime}-\bar{\Sigma}_{5^\prime})/\bar{\Sigma}_{5^\prime}$,
$\Sigma_{5\arcmin}$ and $\bar{\Sigma}_{5\arcmin}$ being the local and
mean surface density measured in circular cells of 5\arcmin\ radius.
Negative contours representing underdense regions are indicated by
blue lines, while positive contours representing overdense regions are
indicated by red lines. The green dashed lines indicate the mean
density.  The distribution of proto-cluster galaxies (large circles)
correlates strongly with positive enhancements in the local
$i$-dropout density distribution, indicating that these are the sites
of formation of some of the first clusters. 
In Fig. \ref{fig:npointing} we plot the frequency distribution of the
$i$-dropouts shown in Fig. \ref{fig:field}, based on a counts-in-cells
analysis of 20,000 randomly placed ACS-sized fields of
$3.4\arcmin\times3.4\arcmin$ (solid histograms). On average, one
expects to find about 2 $i$-dropouts in a random ACS pointing down to
a depth of \zp$=$26.5, but the distribution is skewed with respect to
a pure Poissonian distribution as expected due to the effects of
gravitational clustering. The Poissonian expectation for a mean of 2
$i$-dropouts is indicated by a thin line for comparison. The panel on
the right shows a zoomed-in view to give a better sense of the small
fraction of pointings having large numbers of $i$-dropouts. Also in
Fig. \ref{fig:npointing} we have indicated the counts histogram
derived from a similar analysis performed on $i$-dropouts extracted
from the GOODS survey using the samples of B06. The GOODS result is
indicated by the dotted histogram, showing that it lies much closer to
the Poisson expectation than the MR mock survey. This is of course
expected as our mock survey covers an area over 200$\times$\ larger than
GOODS and includes a much wider range of environments. To illustrate
that the (small) fraction of pointings with the largest number of
objects is largely due to the presence of regions associated with
proto-clusters, we effectively ``disrupt'' all protoclusters by
randomizing the positions of all protocluster galaxies and repeat the
counts-in-cells calculation. The result is shown by the dashed histograms in
Fig. \ref{fig:npointing}. The excess counts have largely disappeared,
indicating that they were indeed due to the proto-clusters. The counts
still show a small excess over the Poissonian distribution due to the
overall angular clustering of the $i$-dropout population.
  
\begin{figure}
\begin{center}
\includegraphics[width=\columnwidth]{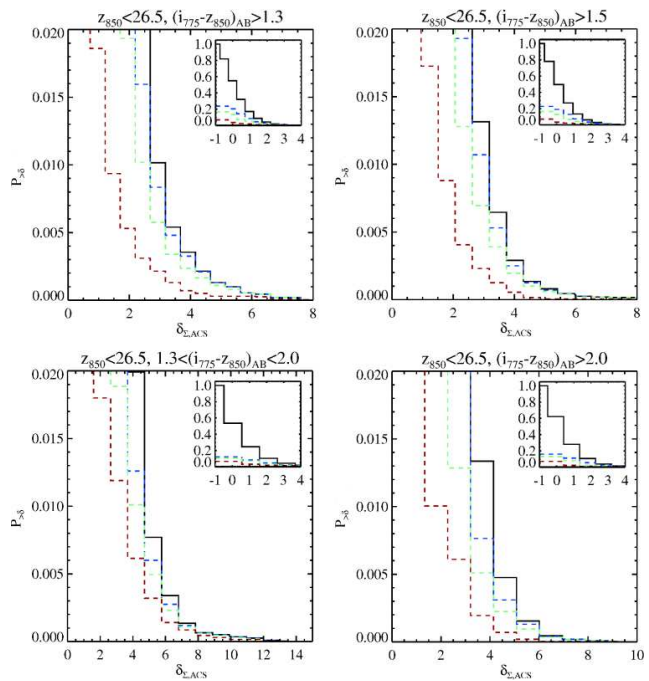}
\end{center}
\caption{\label{fig:pcumul}Panels show the cumulative probability
  distributions of finding regions having a surface overdensity
  $>\delta_{\Sigma,ACS}$ of $i$-dropouts for the four samples extracted
  from the MR mock survey based on colour cuts of \ip--\zp$>1.3$
  (top-left), \ip--\zp$>1.5$ (top-right), $1.3<$\ip--\zp$<2.0$
  (bottom-left), and \ip--\zp$>2.0$ (bottom-right). The inset plots
  show the full probability distributions. Dashed, coloured lines
  indicate the joint probability of finding cells having an
  overdensity $>\delta_{\Sigma,ACS}$ and those cells consisting of at
  least 25\% (blue), 50\% (green) and 75\% proto-cluster galaxies.}
\end{figure}

We can use our counts-in-cells analysis to predict the cumulative
probability, $P_{>\delta}$, of randomly finding an $i$-dropout
overdensity equal or larger than $\delta_{\Sigma,ACS}$. The resuls are
shown in Fig. \ref{fig:pcumul}. The four panels correspond to the
subsamples defined using the four different \ip--\zp\ colour cuts (see
\S2.5 and Fig. \ref{fig:nz2}). Panel insets show the full probability
range for reference. The figure shows that the probability of finding,
for example, cells having a surface overdensity of $i$-dropouts of
$\gtrsim$3 is about half a percent for the \ip--\zp$>$1.3 samples (top
left panel, solid line). The other panels show the dependence of
$P_{>\delta}$ on $i$-dropout samples selected using different colour
cuts. As the relative contribution from for- and background galaxies
changes, the density contrast between real, physical overdensities on
small scales and the ``field'' is increased. 

\begin{figure}
\begin{center}
\includegraphics[width=0.7\columnwidth]{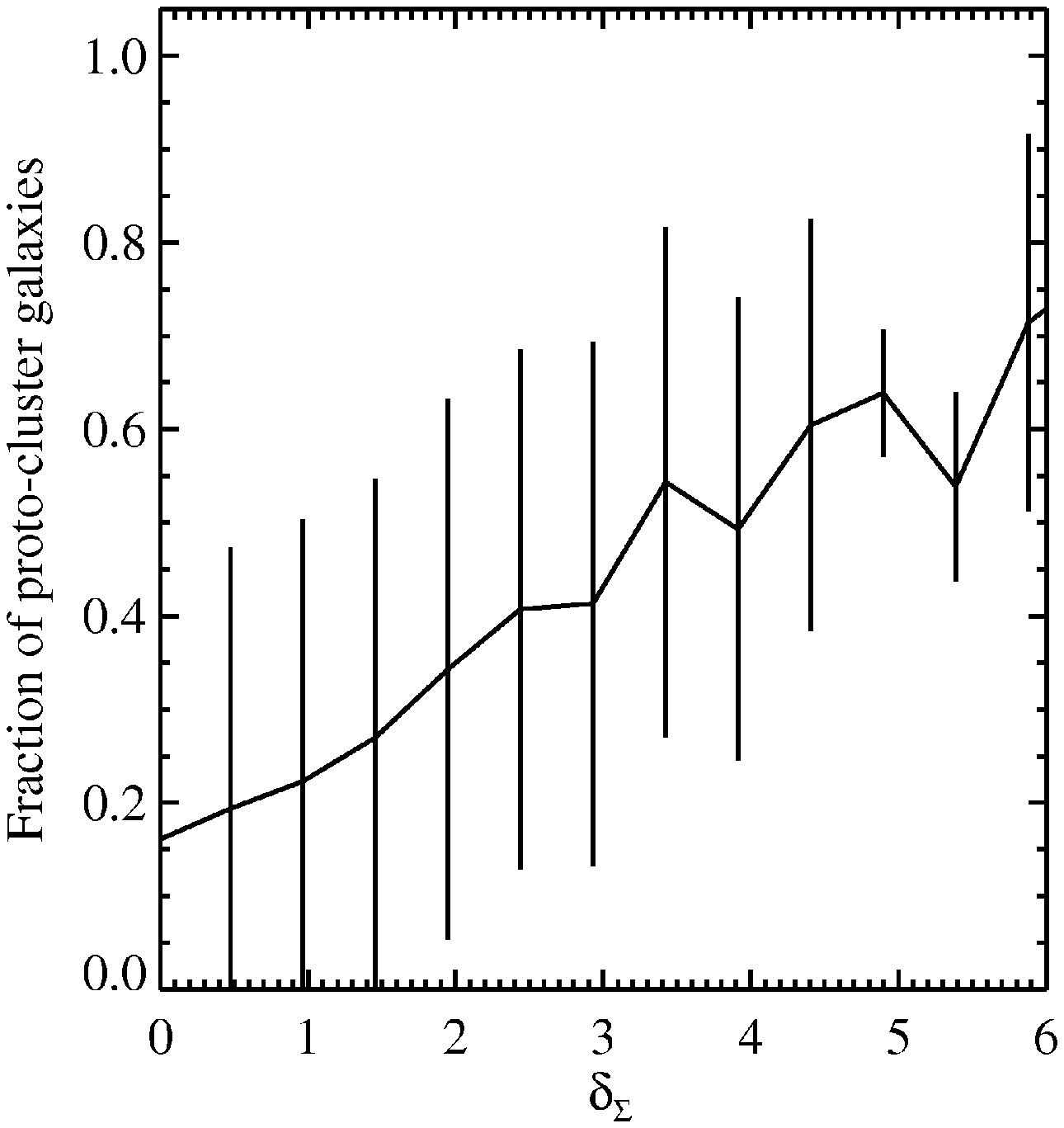}
\end{center}
\caption{\label{fig:deltaclusfrac}The average fraction of
  $i$-dropouts marked as proto-cluster galaxies contained in ACS-sized
  cells as a function of cell overdensity. Error bars are
  $1\sigma$. There is a clear trend showing that larger surface overdensities
  are associated with a larger contribution from galaxies in
  proto-clusters, albeit with significant scatter.}
\end{figure}

The results presented in
Fig. \ref{fig:pcumul} provide us with a powerful way to interpret 
many observational findings. Specifically, overdensities of
$i$-dropouts have been interpreted as evidence for large-scale
structure associated with proto-clusters, at least qualitatively.
Although Fig. \ref{fig:pcumul} tells us the likelihood of finding a
given overdensity, this is not sufficient by itself to answer the
question whether that overdensity is related to a proto-cluster due to
a combination of several effects. First, because we are mainly working
with photometrically selected samples consisting of galaxies spanning
about one unit redshift, projection effects are bound to give rise to
a range of surface densities. Second, the number counts may show
significant variations as a function of position and environment
resulting from the large-scale structure. The uncertainties in the cosmic variance can be reduced
by observing fields that are larger than the typical scale length of
the large-scale structures, but this is often not achieved in typical
observations at $z\sim6$. Third, surface overdensities that are
related to genuine overdensities in physical coordinates are not
necessarily due to proto-clusters, as we have shown that the
descendents of $i$-dropouts can be found in a wide range of
environments at $z=0$, galaxy groups being the most common (see
Fig. \ref{fig:z0halos}). We have separated the contribution of these
effects to $P_{>\delta}$ from that due to proto-clusters by
calculating the fraction of actual proto-cluster $i$-dropouts in each
cell of given overdensity $\delta$. The results are also shown in
Fig. \ref{fig:pcumul}, where dashed histograms indicate the combined
probability of finding a cell of overdensity $\ge\delta$ consisting of
more than 25 (blue lines), 50 (green lines) and 75\% (red lines)
protocluster galaxies. The results show that while, for example, the
chance of $P(\delta\ge2.5)$ is about 1\%, the chance that at least
50\% of the galaxies in such cells are proto-cluster galaxies is only
half that, about 0.5\% (see top left panel in
Fig. \ref{fig:pcumul}). The figure goes on to show that the fractions
of protocluster galaxies increases significantly as the overdensity
increases, indicating that the largest (and rarest) overdensities in
the $i$-dropout distribution are related to the richest proto-cluster
regions. This is further illustrated in Fig. \ref{fig:deltaclusfrac} in which
we plot the average and scatter of the fraction of proto-cluster
galaxies as a function of $\delta$. Although the fraction rises as the
overdensity increases, there is a very large scatter. At
$\delta\approx4$ the average fraction of protocluster galaxies is
about 0.5, but varies significantly between 0.25 and 0.75 (1$\sigma$).
\begin{figure*}
\begin{center}
\includegraphics[width=0.8\textwidth]{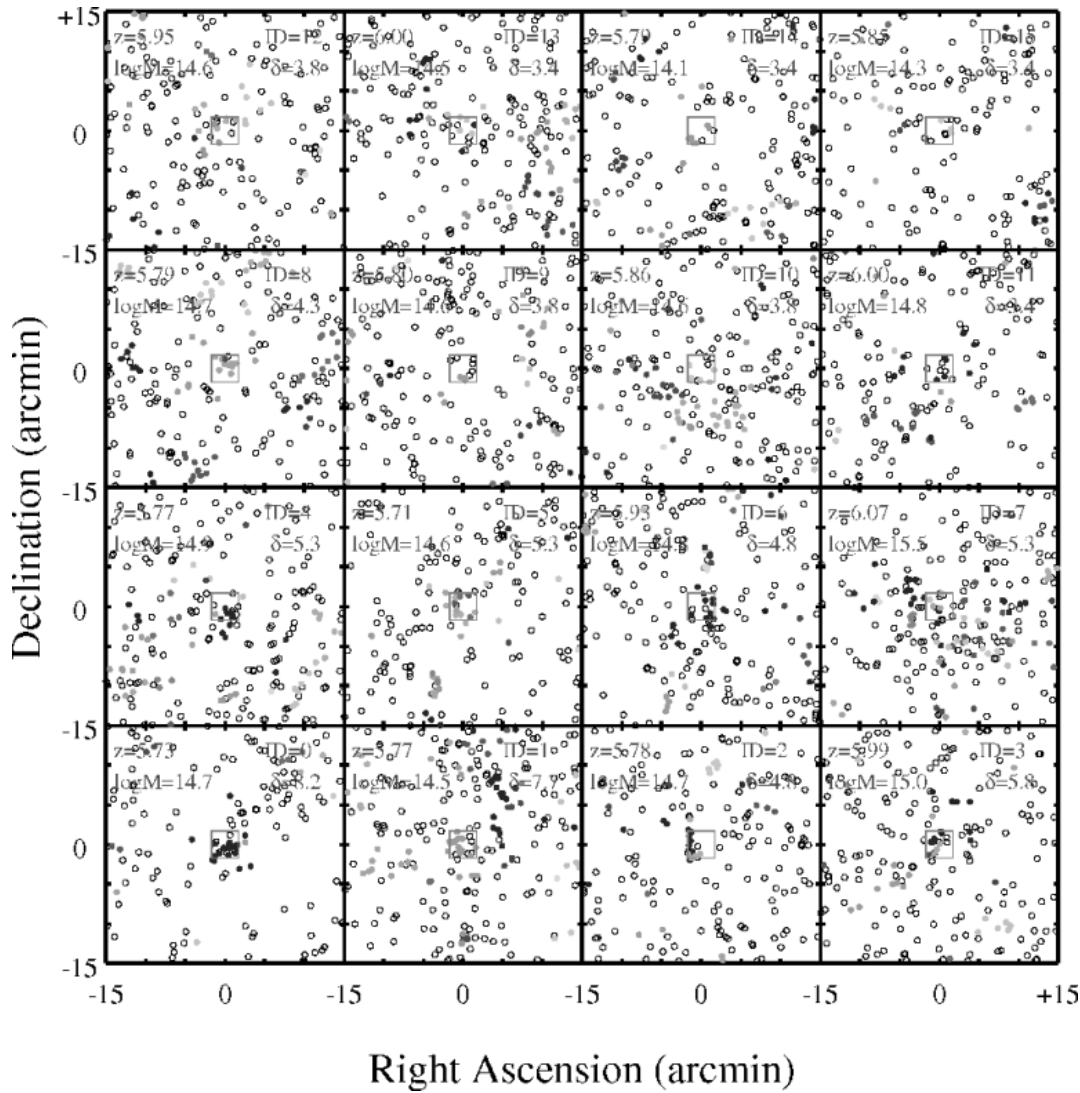}
\end{center}
\caption{\label{fig:topclusters_radec}Panels show the angular
  distribution of \ip-dropouts in 30\arcmin$\times$30\arcmin\ areas
  centered on each of 16 protoclusters associated with overdensities
  $\delta_{\Sigma,ACS}\gtrsim3$. Field galaxies are drawn as open
  circles, and cluster galaxies as filled circles that are colour
  coded according to their cluster membership. The ACS field-of-view
  is indicated by a red square. Numbers near the top of each panel
  indicate the ID, redshift, overdensity and cluster mass (at $z=0$)
  of the protoclusters in the center of each panel.}
\end{figure*}

It will be virtually impossible to estimate an accurate cluster mass at $z=0$ from
a measured surface overdensity at $z\sim6$. Although there
is a correlation between cluster mass at $z=0$ and $i$-dropout
overdensity at $z\sim6$, the scatter is significant. Many of the most
massive ($\cal{M}$$>$$10^{15}$ $M_\odot$) clusters have very small
associated overdensities, while the progenitors of fairly low mass
clusters ($\cal{M}$$\sim$$10^{14}$ $M_\odot$) can be found associated
with regions of relatively large overdensities. However, the 
largest overdensities are consistently associated with the progenitors
of $\cal{M}$$\sim5\times10^{14}-1\times10^{15}$ $M_\odot$ clusters.

\begin{figure}
\begin{center}
\includegraphics[width=\columnwidth]{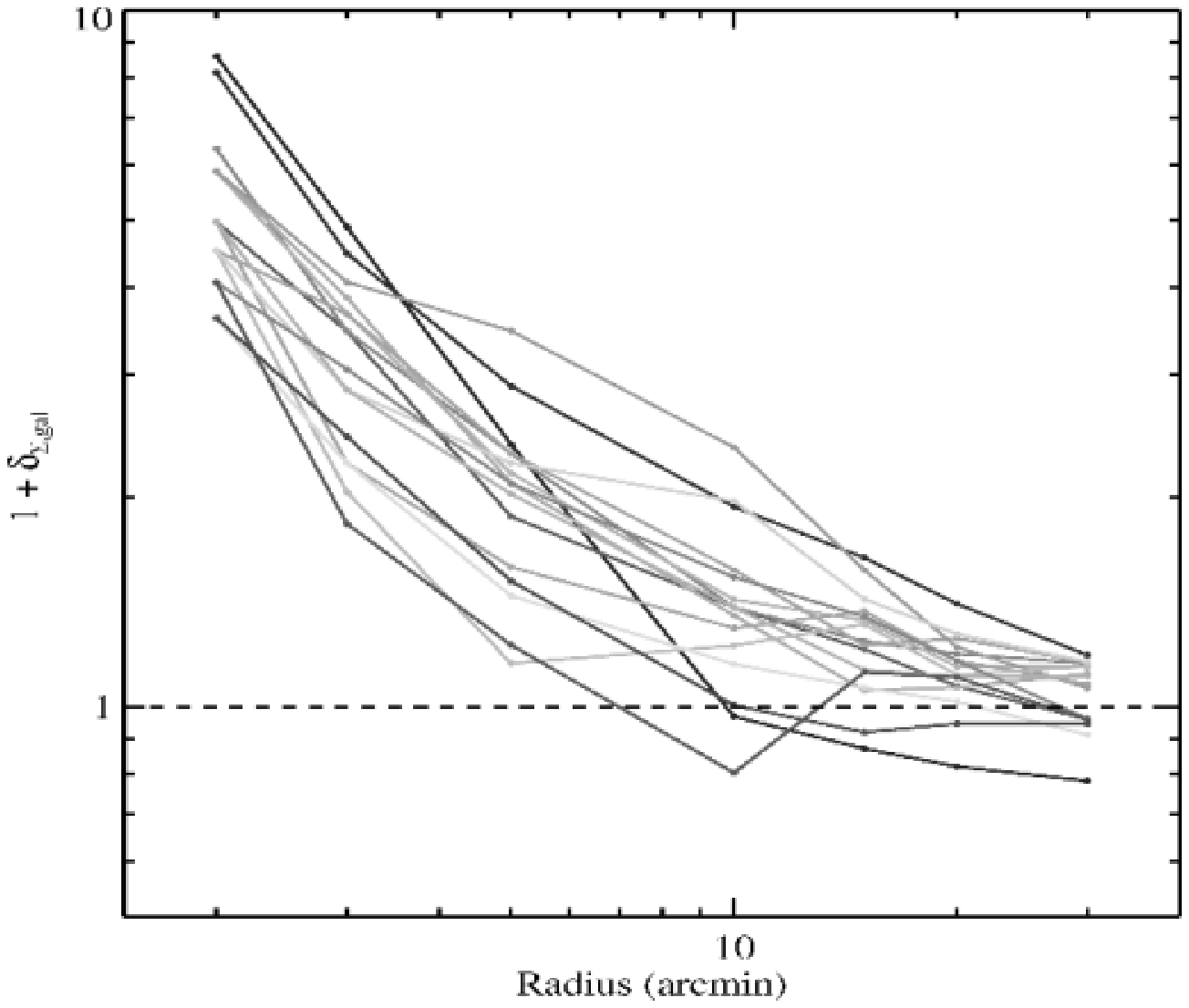}
\end{center}
\caption{\label{fig:topclusters_radius}Lines show overdensity 
 as a function of radius for each of the protocluster
  regions shown in Fig. \ref{fig:topclusters_radec}.}
\end{figure}

\subsection{Some examples}

\begin{figure*}
\begin{center}
\includegraphics[width=0.8\textwidth]{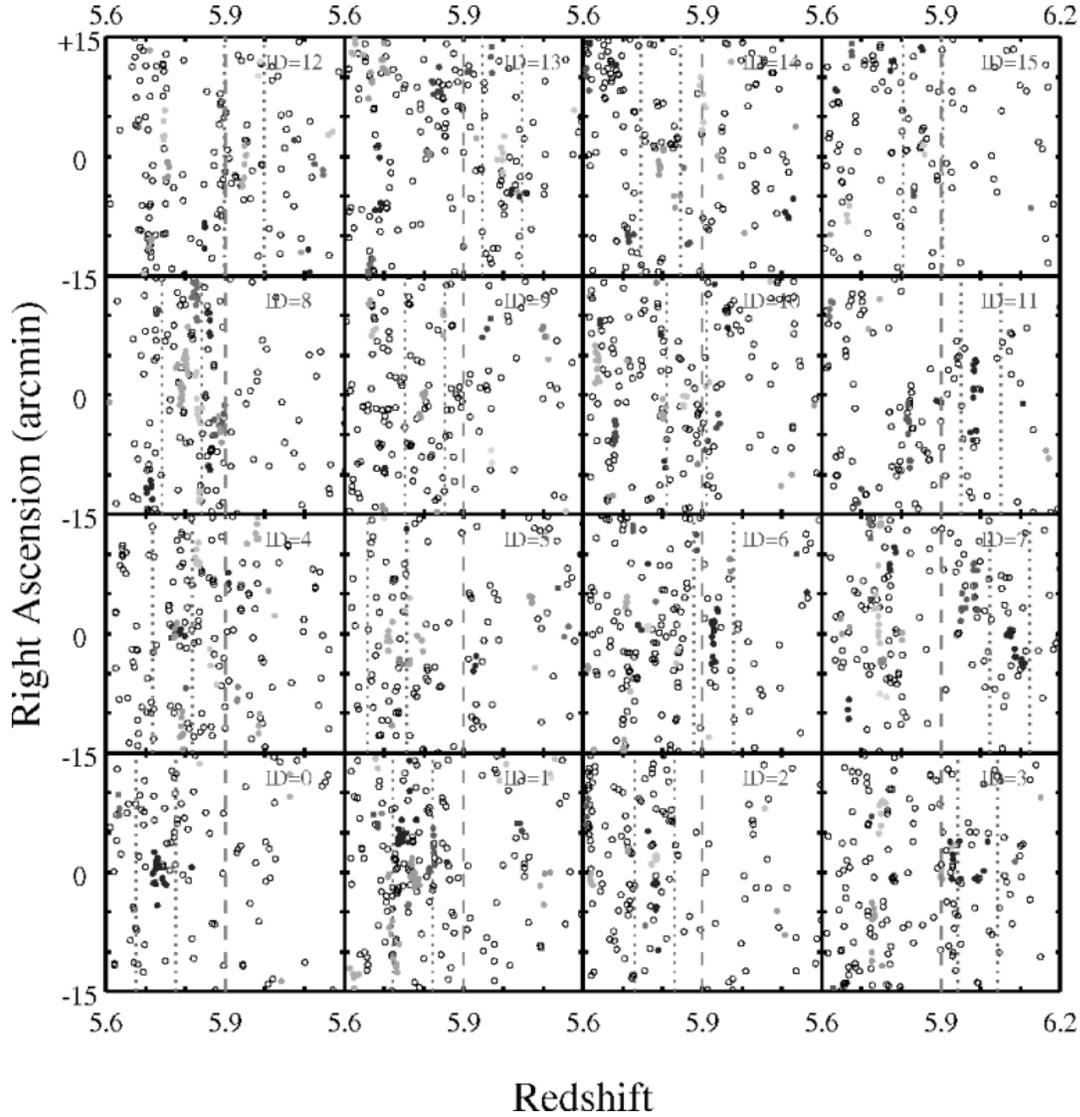}
\end{center}
\caption{\label{fig:topclusters_z}Panels show redshift versus one of
  the 
  angular coordinates of \ip-dropouts for each of the protocluster
  regions shown in Fig. \ref{fig:topclusters_radec}. Field galaxies
  are drawn as open circles, and cluster galaxies as filled circles
  that are  colour coded according to their cluster membership as in
  Fig. \ref{fig:topclusters_radec}. Red dashed lines mark $z=5.9$,
  which roughly corresponds to, respectively, the upper and lower
  redshift of samples selected by placing a cut at \ip--\zp$\lesssim$2
  and \ip--\zp$\gtrsim$2. Blue dotted lines mark the redshift range ($\Delta
  z\approx0.1$) probed by narrowband \lya\ filters.}
\end{figure*}

\noindent
Although the above sections yield useful statistical results, it is
interesting to look at the detailed angular and redshift distributions of the
\ip-dropouts in a few of the overdense regions. In
Fig. \ref{fig:topclusters_radec} we show 16 30\arcmin$\times$30\arcmin\
regions having overdensities ranging from $\delta_{\Sigma,ACS}\sim8$ (bottom left
panel) to $\sim3$ (top right panel). In each panel we indicate the
relative size of an ACS pointing (red square), and the redshift,
overdensity and present-day mass of the most massive protoclusters 
are given in the
top left and right corners. Field galaxies are drawn as open circles,
while protocluster galaxies are drawn as filled circles. Galaxies
belonging to the same proto-cluster are drawn in the same
colour. While some regions contain relatively compact protoclusters
with numerous members inside the 3.4\arcmin$\times$3.4\arcmin\ ACS
field-of-view (e.g. panels \#0, 1 and 8), other regions may contain
very few or highly dispersed galaxies. Also, many regions contain
several overlapping protoclusters as the selection function is
sensitive to structures from a relatively wide range in redshift
inside the 30\arcmin$\times$30\arcmin\ regions plotted.  Although the
angular separation between galaxies belonging to the same protocluster
is typically smaller than $\sim$10\arcmin\ or 25 Mpc (comoving),
Fig. \ref{fig:topclusters_radius} shows that the overdensities of
regions centered on the protoclusters are significantly positive out
to much larger radii of between 10 to 30\arcmin, indicating that the
protoclusters form inside very large filaments of up to 100 Mpc in
size that contribute significantly to the overall (field) number
counts in the protocluster regions. In Fig. \ref{fig:topclusters_z} we
plot the redshift coordinate against one of the angular coordinates
using the same regions and colour codings as in
Fig. \ref{fig:topclusters_radec}. Protoclusters are significantly more
clumped in redshift space compared to field galaxies, due to
flattening of the velocity field associated with the collapse of large
structures. In each panel, a red dashed line marks $z=5.9$, which
roughly corresponds to, respectively, the upper and lower redshift of
samples selected by placing a cut at \ip--\zp$\lesssim$2 and
\ip--\zp$\gtrsim$2 (see the redshift selection functions in
Fig. \ref{fig:nz2}). Such colour cuts may help reduce the contribution
from field galaxies by about 50\%, depending on the redshift one is
interested in. We also mark the typical redshift range of $\Delta
z\approx0.1$ probed by narrowband filters centered on the redshift of
each protocluster using blue dotted lines. As we will show in more detail
in \S\ref{sec:ouchi} below, such narrowband selections
offer one of the most promising methods for finding and studying the
earliest collapsing structures at high redshift, because of the  
significant increase in
contrast between cluster and field regions. However, such surveys
are time-consuming and only probe the part of the galaxy population that
is bright in the \lya\ line.

\section{Comparison with observations from the literature} 

\noindent
Our mock survey of $i$-dropouts constructed from the MR, due to its
large effective volume, spans a wide range of environments and is
therefore ideal for making detailed comparisons with observational
studies of the large-scale structure at $z\sim6$. In the following
subsections, we will make such comparisons with two studies of 
candidate proto-clusters of $i$-dropouts 
and \lya\ emitters found in the SDF and SXDF.

\subsection{The candidate proto-cluster of \citet{ota08}}

\noindent
When analysing the sky distribution of $i$-dropouts in the 876
arcmin$^2$ Subaru Deep Field, \citet{ota08} (henceforward O08) discovered a large
surface overdensity, presumed to be a proto-cluster at $z\sim6$. The
magnitude of the overdensity was quantified as the excess of
$i$-dropouts in a circle of 20 Mpc comoving radius. The 
region had $\delta_{\Sigma,20\mathrm{Mpc}}=0.63$ with $3\sigma$
significance. Furthermore, this region also contained the highest
density contrast measured in a 8 Mpc comoving radius
$\delta_{\Sigma,8\mathrm{Mpc}}=3.6$ ($5\sigma$) compared to other
regions of the SDF. By relating the total overdensity in dark matter
to the measured overdensity in galaxies through an estimate of the
galaxy bias parameter, the authors estimated a mass for the
proto-cluster region of $\sim1\times10^{15}$ $M_\odot$.

\begin{figure}
\begin{center}
\includegraphics[width=\columnwidth]{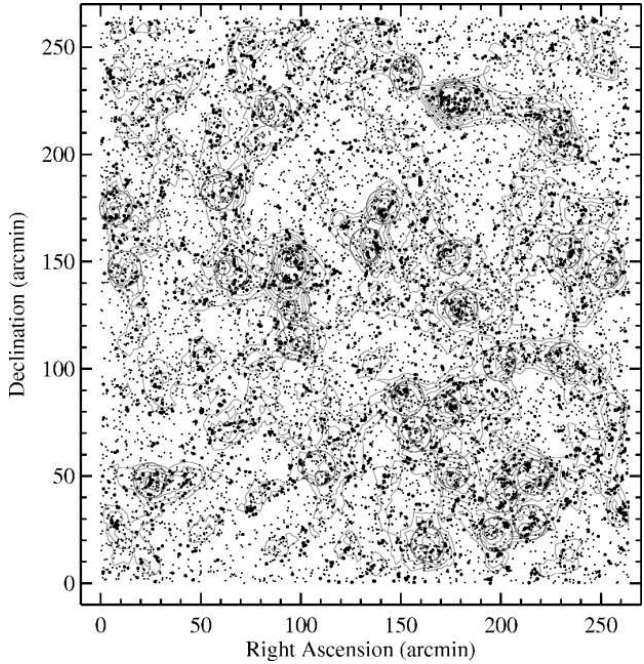}
\end{center}
\caption{\label{fig:ota}The sky distribution of $i$-dropouts selected
  using criteria matched to those of \citet{ota08}. Grey solid lines
  are surface density contours of
  $\delta_{\Sigma,20\mathrm{Mpc}}=0,+0.2,+0.4,+0.6,+0.8$ and
  $+1.0$. Large red dashed circles mark overdense regions of
  $\delta_{\Sigma,20\mathrm{Mpc}}>0.63$, corresponding to similar
  overdensities as that associated with the candidate $z\sim6$
  proto-cluster region found by \citet{ota08} in the Subaru Deep
  Field.  Small red circles inside each region mark a subregion having
  the largest overdensity $\delta_{\Sigma,8\mathrm{Mpc}}$ measured in
  a 8 Mpc co-moving radius (projected) cell (see text for further details).}
\end{figure}

We use our mock survey to select $i$-dropouts with \ip--\zp$>1.5$ and
\zp$<26.5$, similar to O08. The resulting surface density was 0.16
arcmin$^{-2}$ in very good agreement with the value of 0.18
arcmin$^{-2}$ found by O08. In Fig. \ref{fig:ota} we plot the sky
distribution of our sample, and connect regions of constant (positive)
density $\delta_{\Sigma,20\mathrm{Mpc}}$.  Next we selected all
regions that had $\delta_{\Sigma,20\mathrm{Mpc}}\ge0.63$. These
regions are indicated by the large red circles in
Fig. \ref{fig:ota}. We find $\sim$30 (non-overlapping) regions in our
entire mock survey having $\delta_{\Sigma,20\mathrm{Mpc}}=0.6-2.0$ at
$2-7\sigma$ significance, relative to the mean dropout density of
$\bar{\Sigma}_{20\mathrm{Mpc}}\approx32$.  Analogous to
Fig. \ref{fig:field}, we have marked all $i$-dropouts associated with
proto-clusters with large symbols. It can be seen clearly that the
proto-cluster galaxies are found almost exclusively inside the regions
of enhanced local surface density indicated by the contour lines,
while the large void regions are virtually depleted of proto-cluster
galaxies.  Although the 30 regions of highest overdensity selected to
be similar to the region found by O08 coincide with the highest peaks
in the global density distribution across the field, it is interesting
to point out that in some cases the regions contain very few actual
proto-cluster galaxies, e.g., the regions at (RA,DEC)=(10,150) and (80,220) in
Fig. \ref{fig:ota}. We therefore introduce a proto-cluster ``purity''
parameter, $\cal{R}_{\mathrm{pc}}$, defined as the ratio of galaxies
in a (projected) region that belong to protoclusters to the total
number of galaxies in that region.  We find
$\cal{R}_{\mathrm{pc,20Mpc}}\approx$16--50\%. The purest or richest
proto-clusters are found in regions having a wide range in
overdensities, e.g., the region at (175,225) with
$\delta_{\Sigma,20\mathrm{Mpc}}=2.2$,
$\cal{R}_{\mathrm{pc,20Mpc}}=$50\%), and the region at (200,40) with
$\delta_{\Sigma,20\mathrm{Mpc}}=0.9$,
$\cal{R}_{\mathrm{pc,20Mpc}}=$40\%.  Following O08 we also calculate
the maximum overdensity in each region using cells of 8 Mpc radius. We
find $\delta_{\Sigma,8\mathrm{Mpc}}=1.1-3.5$ with $2-6\sigma$
significance. These sub-regions are indicated in Fig. \ref{fig:ota}
using smaller circles. Interestingly, there is a very wide range in
proto-cluster purity of
$\cal{R}_{\mathrm{pc,8Mpc}}\approx$0--80\%. The largest overdensity in
Fig. \ref{fig:ota} at (175,225) corresponds to the region giving birth
to the most massive cluster. By $z=0$, this region has grown into a
``supercluster'' region containing numerous clusters, two of which have
$M>10^{15}$ $M_\odot$.

We conclude that local overdensities in the distribution of
$i$-dropouts on scales of $\sim$10-50 comoving Mpc similar to the one
found by O08 indeed trace the seeds of massive clusters. Because our mock
survey is about 80$\times$\ larger than the SDF, we expect that one
would encounter such proto-cluster regions in about one in three (2.7)
SDF-sized fields on average. However, the fraction of actual
proto-cluster galaxies is in the range 16--50\% (0--80\% for 8 Mpc
radius regions). This implies that while one can indeed find the
overdense regions where clusters are likely to form, there is no way
of verifying which galaxies are part of the proto-cluster and which
are not, at least not when using photometrically selected samples.
These results are consistent with our earlier finding that there is a
large scatter in the relation between the measured surface overdensity
and both cluster ``purity'' and the mass of its descendant cluster at $z=0$ (Sect. \ref{sec:detclus}). 

\subsection{The \lya-selected proto-cluster of \citet{ouchi05}}
\label{sec:ouchi}

\begin{figure}
\begin{center}
\includegraphics[width=\columnwidth]{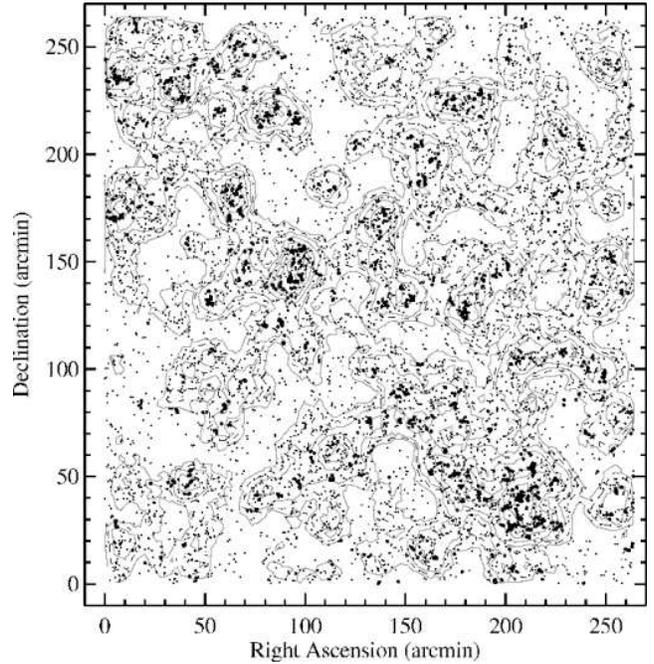}
\end{center}
\caption{\label{fig:ouchi}Mock \lya\ survey at $\simeq5.8\pm0.05$
  constructed from the MR mock \ip-dropout sample. Grey solid lines
  are surface density contours of
  $\delta_{\Sigma,20\mathrm{Mpc}}=-0.25$ to 3.25 with a step increase
  of 0.5 as in Fig. 2 of \citet{ouchi05}. The black dashed line marks
  the average field density. Small circles indicate field
  galaxies. Large circles indicate protocluster galaxies.}
\end{figure}

\noindent
The addition of velocity information gives studies of \lya\ samples a
powerful edge over purely photometrically selected \ip-dropout
samples.  As explained by \citet[][and references therein]{monaco05},
peculiar velocity fields are influenced by the large-scale structure:
streaming motions can shift the overall distribution in redshift,
while the dispersion can both increase and decrease as a result of
velocity gradients. Galaxies located in different structures that are
not physically bound will have higher velocity dispersions, while
galaxies that are in the process of falling together to form
non-linear structures such as a filaments, sheets (or ``pancakes'')
and proto-clusters will have lower velocity dispersions.

Using deep narrow-band imaging observations of the SXDF,
\citet{ouchi05} (O05) were able to select \lya\ candidate galaxies at
$z\simeq5.7\pm0.05$. Follow-up spectroscopy of the candidates in one
region that was found to be significantly overdense ($\delta\gtrsim3$)
on a scale of 8 Mpc (comoving) radius resulted in the discovery of two
groups ('A' and 'B') of \lya\ emitting galaxies each having a very
narrow velocity dispersion of $\lesssim200$ km s$^{-1}$. The
three-dimensional density contrast is on the order of $\sim100$,
comparable to that of present-day clusters, and the space density of
such proto-cluster regions is roughly consistent with that of massive
clusters (see O05).

In order to study the velocity fields of collapsing structures and
carry out a direct comparison with O05, we construct a simple
\lya\ survey from our mock sample as follows.  First, we construct a
(Gaussian) redshift selection function centred on $z=5.8$ with a
standard deviation of 0.04. As it is not known what causes some
galaxies to be bright in \lya\ and others not, our simulations do
not include a physical prescription for \lya\ as such. However,
empirical results suggest that \lya\ emitters are mostly young,
relatively dust-free objects and a subset of the \ip-dropout
population.  The fraction of galaxies with high equivalent width
\lya\ is about 40\%, and this fraction is found to be roughly constant
as a function of the rest-frame UV continuum magnitude. Therefore, we
scale our selection function so that it has a peak selection
efficiency of 40\%.  Next, we apply this selection function to the
\ip-dropouts from the mock survey to create a sample with a 
redshift distribution similar to that expected from a narrowband
\lya\ survey. Finally, we tune the limiting \zp\ magnitude until we
find a number density that is similar to that reported by O05. By
setting \zp$<$26.9 mag we get the desired number density of $\sim$0.1
arcmin$^{-2}$. The mock \lya\ field is shown in Fig. \ref{fig:ouchi}.

\begin{figure}
\begin{center}
\includegraphics[width=\columnwidth]{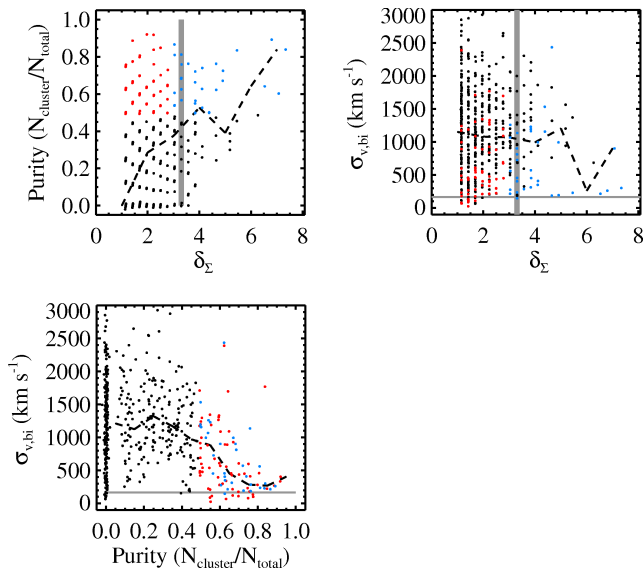}
\end{center}
\caption{\label{fig:ouchiresults}The correlations between surface
  overdensity, cluster purity and velocity dispersion for
  \lya\ galaxies selected from the mock \lya\ survey shown in
  Fig. \ref{fig:ouchi} using randomly drawn cells of 8 Mpc (comoving)
  radius. Dashed lines indicate the median trends. Red points
  highlight regions of purity $\cal{R}>$0.5. Blue points highlight
  regions of $\cal{R}>$0.5 and $\delta_\Sigma>3$. Shaded areas mark the
  values obtained by \citet{ouchi05} for a protocluster of
  \lya\ galaxies in the SDF. See text for details. }
\end{figure}

In the top left panel of Fig. \ref{fig:ouchiresults} we plot the
overdensities measured in randomly drawn regions of 8 Mpc (comoving)
radius against the protocluster purity parameter, analogous to
Fig. \ref{fig:deltaclusfrac}. Although the median purity of a sample
increases with overdensity (dashed line), the scatter indicated by the
points is very large even for overdensities as large as
$\delta\approx3$ found by O05 (marked by the shaded region in the top
panel of Fig. \ref{fig:ouchiresults}). To guide the eye, we have
plotted regions of purity $>$0.5 as red points, and regions having
purity $>$0.5 and $\delta>3$ as blue points in all panels of
Fig. \ref{fig:ouchiresults}. Next, we calculate the velocity
dispersion, $\sigma_{v,bi}$, from the peculiar velocities of the
galaxies in each region using the bi-weight estimator of
\citet{beers90} that is robust for relatively small numbers of objects
($N\simeq10-50$), and plot the result against $\delta$ and cluster
purity in the top right and bottom left panels of
Fig. \ref{fig:ouchiresults}, respectively.  

Although gravitational clumping of galaxies in redshift space causes
the velocity dispersions to be considerably lower than the velocity
width allowed by the bandpass of the narrowband filter
($\langle\sigma_{v,bi}\rangle\simeq1000$ km $s^{-1}$ compared to
$\sigma_{NB}\approx1800$ km $s^{-1}$ for $\sigma_{NB,z}=0.04$), the
velocity dispersion is not a decreasing function of the overdensity
(at least not up to $\delta\approx3-4$) and the scatter is
significant. This can be explained by the fact that proto-clusters
regions are rare, and even regions that are relatively overdense in
angular space still contain many galaxies that are not contained
within a single bound structure.  A
much stronger correlation is found between dispersion and cluster
purity (see bottom left panel of
Fig. \ref{fig:ouchiresults}). Although the scatter in dispersion is
large for regions with a purity of $\lesssim0.5$, the smallest
dispersions are associated with some of the richest protocluster
regions. This can be understood because the ``purest'' structures
represent the bound inner cores of future clusters at $z=0$. The
velocity dispersions are low because these systems do not contain many
field galaxies that act to inflate the velocity dispersion
measurements. 
Therefore, the velocity dispersion correlates much more
strongly with the protocluster purity than with the surface
overdensity. The overdensity parameter helps, however, in reducing
some of the ambiguity in the cluster richness at small dispersions
(compare black and blue points at small $\sigma_{v,bi}$ in the bottom
left panel). The shaded regions in Fig. \ref{fig:ouchiresults}
indicate the range of measurements of O05, implying that their structure has
the characteristics of \lya\ galaxies falling into a protocluster at
$z\sim6$.

\section{Where is the large-scale structure associated with $z\sim6$ QSOs?}
\label{sec:qso}

\noindent
For reasons explained in the Introduction, it is generally assumed
that the luminous QSOs at $z\sim6$ inhabit the most
massive dark matter in the early Universe. The HST/ACS, with its deep and
wide-field imaging capabilities in the \ip\ and \zp\ bands, has made
it possible to test one possible implication of this by searching for
small neighbouring galaxies tracing these massive halos. In this Section, we will first
investigate what new constraint we can put on the masses of the host
halos based on the observed neighbour
statistics. \citet{munoz08a} have addressed the same problem based on 
the excursion set formalism. Our analysis 
is based on semi-analytic models incorporated in the MR simulation,
which we believe is likely to provide a more realistic description of
galaxy properties at $z\sim6$.  
We will use the simulations to evaluate what we can say about the most likely
environment of the QSOs and whether they are associated with
proto-clusters.  We finish the Section by presenting some clear
examples from the simulations that would signal a massive overdensity
in future observations.

Several searches for companion galaxies in the vicinity of $z\sim6$
QSOs have been carried out to date. In Table \ref{tab:surveys} we list
the main surveys, covering in total 6 QSOs spanning the redshift range
$5.8<z<6.4$.
We have used the results given in \citet{stiavelli05}, \citet{zheng06}
and \citet{kim08} to calculate the surface overdensities associated
with each of the QSO fields listed in Table \ref{tab:surveys}. Only
two QSOs were found to be associated with positive overdensities to a
limiting magnitude of \zp$=$26.5: J0836+0054\footnote{The significance
  of the overdensity in this field is less than originally stated by
  \citet{zheng06} as a result of underestimating the contamination
  rate when a \vp\ image is not available to reject lower redshift
  interlopers.} ($z=5.82$) and J1030+0524 ($z=6.28$) both had
$\delta_{\Sigma,\mathrm{ACS}}\approx1$, although evidence suggests
that the overdensity could be as high as $\approx2-3$ when taking into
account subclustering within the ACS field or sources selected using
different $S/N$ or colour cuts \citep[see][for
  details]{stiavelli05,zheng06,ajiki06,kim08}. The remaining four QSO
fields (J1306+0356 at $z=5.99$, J1630+4012 at $z=6.05$, J1048+4637 at
$z=6.23$, and J1148+5251 at $z=6.43$) were all consistent with having
no excess counts with $\delta_{\Sigma,\mathrm{ACS}}$ spanning the
range from about $-$1 to $+$0.5 relatively independent of the method
of selection \citep{kim08}. Focusing on the two overdense QSO fields,
Fig. \ref{fig:pcumul} tells us that overdensities of
$\delta_{\Sigma,\mathrm{ACS}}\gtrsim1$ are fairly common, occurring at
a rate of about 17\% in our $4\degr\times4\degr$ simulation. 
The probability of finding a random field with 
$\delta_{\Sigma}\gtrsim2-3$ is about 5 to 1\%. It is evident that none of the six quasar fields have
highly significant overdensities.  The case for overdensities near the
QSOs would strengthen if all fields showed a systematically higher,
even if moderate, surface density. However, when considering the
sample as a whole the surface densities of $i$-dropouts near $z\sim6$
QSOs are fairly average, given that four of the QSO fields have lower
or similar number counts compared to the field. With the exception
perhaps of the field towards the highest redshift QSO J1148+5251,
which lies at a redshift where the $i$-dropout selection is
particularly inefficient (see Fig. \ref{fig:nz2}), the lack of
evidence for substantial (surface) overdensities in the QSO fields is
puzzling.

\begin{figure}
\begin{center}
\includegraphics[width=\columnwidth]{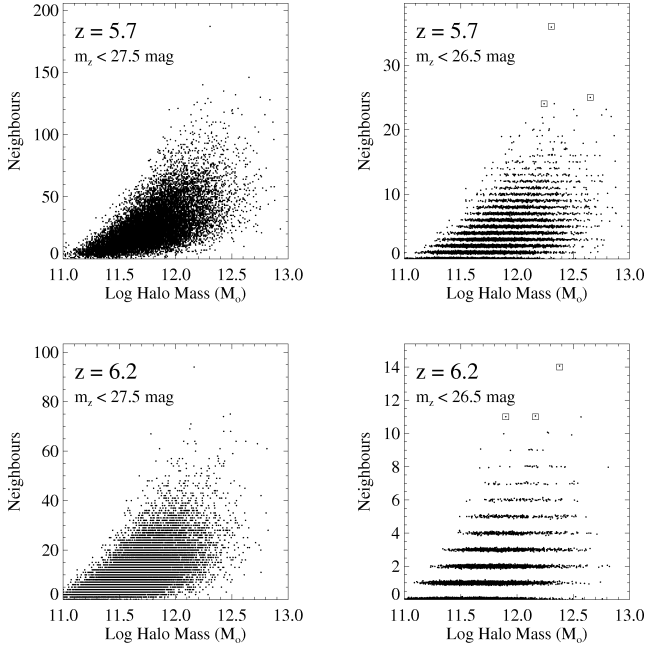}
\end{center}
\caption{\label{fig:boxtest}Panels show the number of neighbours
  (\ip-dropouts) in cubic regions of $(20~h^{-1})^3$ Mpc$^3$ versus
  the mass of the most massive halo found in each of those
  regions. Top and bottom panels are for snapshots at $z=5.7$ and
  $z=6.2$, respectively. Left and right panels are for neighbour
  counts down to limiting magnitudes of \zp=27.5 (left) and \zp$=26.5$
  mag, respectively. There is a wide dispersion in the number of
  neighbours, even for the most massive halos at $z\sim6$. The highest
  numbers of neighbours are exclusively associated with the massive
  end of the halo mass function, allowing one to derive a lower limit
  for the mass of the most massive halo for a given number of
  neighbour counts.  The scatter in the number of neighbours versus
  the mass of the most massive halo reduces signficantly when going to
  fainter magnitudes. The small squares in the panels on the right
  correspond to the three richest regions (in terms of \zp$<$26.5 mag
  dropouts) that are shown in close-up in Fig. \ref{fig:z6panels}.}
\end{figure}

In Fig. \ref{fig:boxtest} we have plotted the number of \ip-dropouts
encountered in cubic regions of $20\times20\times20$ $h^{-1}$ Mpc
against the mass of the most massive dark matter halo found in each
region. Panels on the left and on the right are for limiting
magnitudes of \zp=27.5 and 26.5 mag, respectively. Because the most
massive halos are so rare, here we have used the full MR snapshots at
$z=5.7$ (top panels) and $z=6.2$ (bottom panels) rather than the
lightcone in order to improve the statistics. There is a systematic
increase in the number of neighbours with increasing maximum halo
mass. However, the scatter is very large: for example, focusing on the
neighbour count prediction for $z=5.7$ and \zp$<$26.5 (top right
panel) we see that the number of neighbours of a halo of $10^{12}$
$h^{-1}$ $M_\odot$ can be anywhere between 0 and 20, and some of the
most massive halos of $10^{13}$ $h^{-1}$ $M_\odot$ have a relatively
low number of counts compared to some of the halos of significant
lower mass that are far more numerous.  However, for a given
\zp$<$26.5 neighbour count (in a $20\times20\times20$ $h^{-1}$ Mpc
region) of $\gtrsim5$, the halo mass is {\it always} above
$\sim10^{11.5}$ $h^{-1}$ $M_\odot$, and if one would observe
$\gtrsim25$ \ip-dropout counts one could conclude that that field must
contain a supermassive halo of $\gtrsim10^{12.5}$ $h^{-1}$
$M_\odot$. Thus, in principle, one can only estimate a lower limit on
the maximum halo mass as a function of the neighbour counts. The
left panel shows that the scatter is
much reduced if we are able to count galaxies to a limiting \zp-band
magnitude of 27.5 instead of 26.5, simply because the Poisson noise is
greatly reduced. 

\begin{figure*}
\begin{center}
\includegraphics[width=\textwidth]{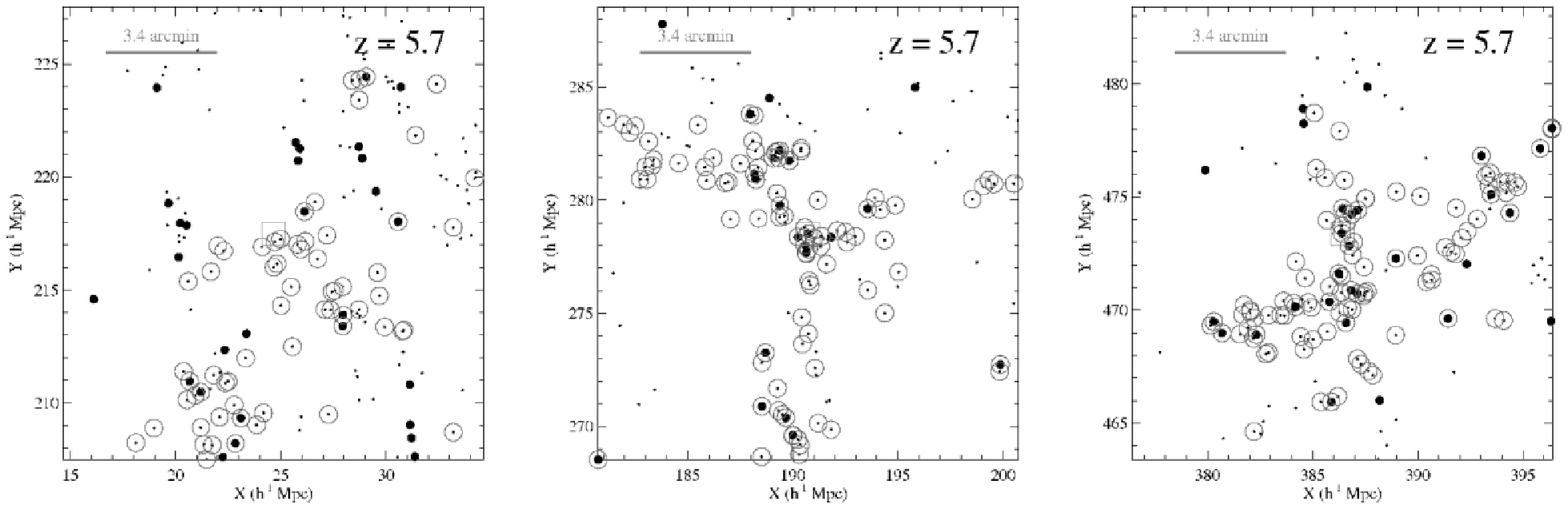}
\includegraphics[width=\textwidth]{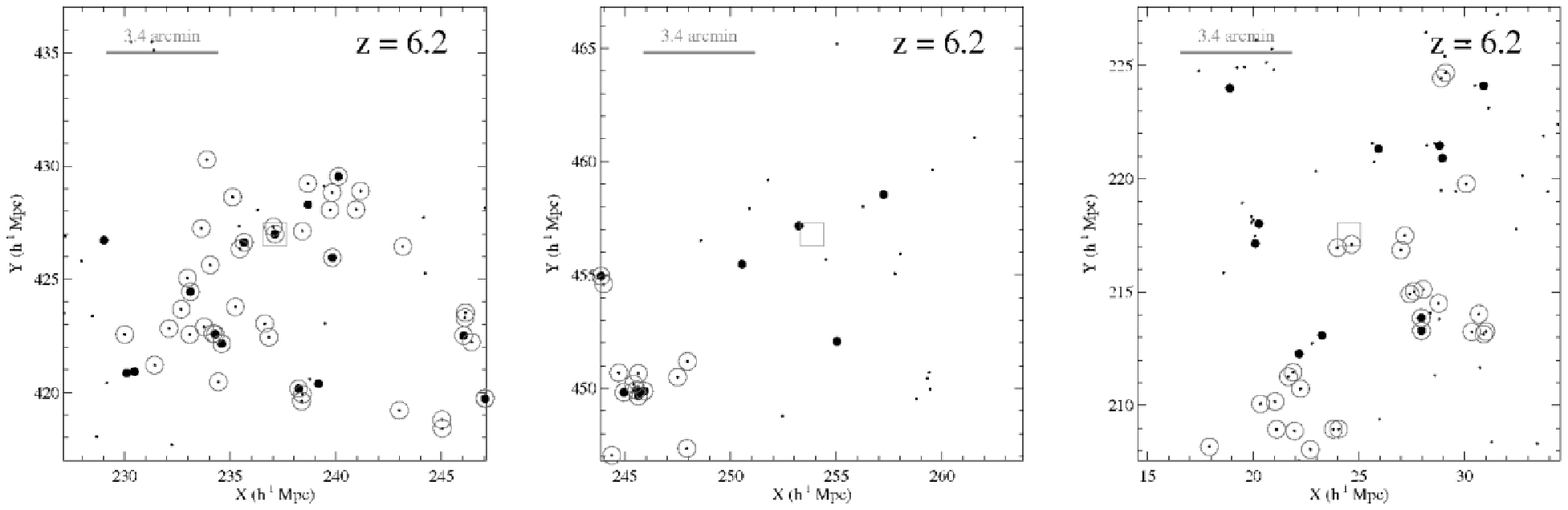}
\end{center}
\caption{\label{fig:z6panels}Close-up views of three $(20~h^{-1})^3$
  Mpc$^3$ regions that were found to be highly overdense in \zp$<$26.5
  mag \ip-dropouts as marked by the squares in
  Fig. \ref{fig:boxtest}. The top row of panels correspond to the
  three richest regions found at $z=5.7$, while the bottom row
  corresponds to those at $z=6.2$. The position of the most massive
  halo in each region is indicated by a green square. Large and small
  dots correspond to dropout galaxies having \zp$<$26.5 and $<$27.5
  mag, respectively. Galaxies that have been identified as part of a
  protocluster structure are indicated by blue circles. The scale bar
  at the top left in each panel corresponds to the size of an ACS/WFC
  pointing used to observe $z\sim6$ QSO fields. Note that the galaxy
  corresponding to the most massive halo as indicated by the green
  square is not always detected in our \ip-dropout survey due to dust
  obscuration associated with very high star formation rates.}
\end{figure*}

We can therefore conclude that the relatively average number of counts
observed in the QSO fields is not inconsistent with the QSOs being
hosted by very massive halos. However, one could make an equally
strong case that they are, in fact, not. If we translate our results
of Fig. \ref{fig:boxtest} to the QSO fields that cover a smaller
projected area of $(\sim5~h^{-1})^2$ Mpc$^2$, and we add back in the
average counts from the fore- and background provided by our lightcone
data, we estimate that for QSOs at $z\approx5.7$ we require an
overdensity of $\delta_{\Sigma,ACS}\sim4$ in order to be able to put a
lower limit on the QSO host mass of $\sim10^{12}$ $M_\odot$, while a
$\delta_{\Sigma,ACS}\sim1$ is consistent with $\sim10^{11}$
$M_\odot$. At $z=6.2$, we would require $\delta_{\Sigma,ACS}\gtrsim2$
for $M\gtrsim10^{12}$ $M_\odot$ and $\delta_{\Sigma,ACS}\gtrsim1$ for
$M\gtrsim10^{11.5}$ $M_\odot$. Comparison with the relatively low
surface overdensities observed thus suggests that the halo mass is
uncontrained by the current data. Nonetheless, we can at least
conclude quite firmly that the QSOs are in far less rich environments
(in terms of galaxy neigbours) compared to many rich regions found
both in the simulations and in some of the deep field surveys
described in the previous Section.  In order to get a feel for what
the QSO fields might have looked like if they were in highly overdense
regions, we present some close-up views in Fig. \ref{fig:z6panels} of
the three richest regions of \zp$<$26.5 mag \ip-dropouts as marked by
the small squares in Fig. \ref{fig:boxtest}. The central position
corresponding to that of the most massive halo in each region is
indicated by the green square.  Large and small dots correspond to
dropout galaxies having \zp$<$26.5 and $<$27.5 mag, respectively. For
reference, we use blue circles to indicate galaxies that have been
identified as part of a protocluster structure. The scale bar at the
top left in each panel corresponds to the size of an ACS/WFC pointing
used to observe $z\sim6$ QSO fields. We make a number of interesting
observations. First, using the current observational limits on depth
(\zp=26.5 mag) and field size (3.4\arcmin, see scale bar) imposed by
the ACS observations of QSOs, it would actually be quite easy to miss
some of these structures as they typically span a larger region of
2--3 ACS fields in diameter. Going too much fainter magnitudes would
help considerably, but this is at present unfeasible. Note, also, that
in three of the panels presented here, the galaxy associated with the
massive central halo does not pass our magnitude limits. It is missed
due to dust obscuration associated with very high star formation rates
inside these halos, implying that they will be missed by large-area UV
searches as well (unless, of course, they also host a luminous,
unobscured QSO).

\begin{figure*}
\begin{center}
\includegraphics[width=\textwidth]{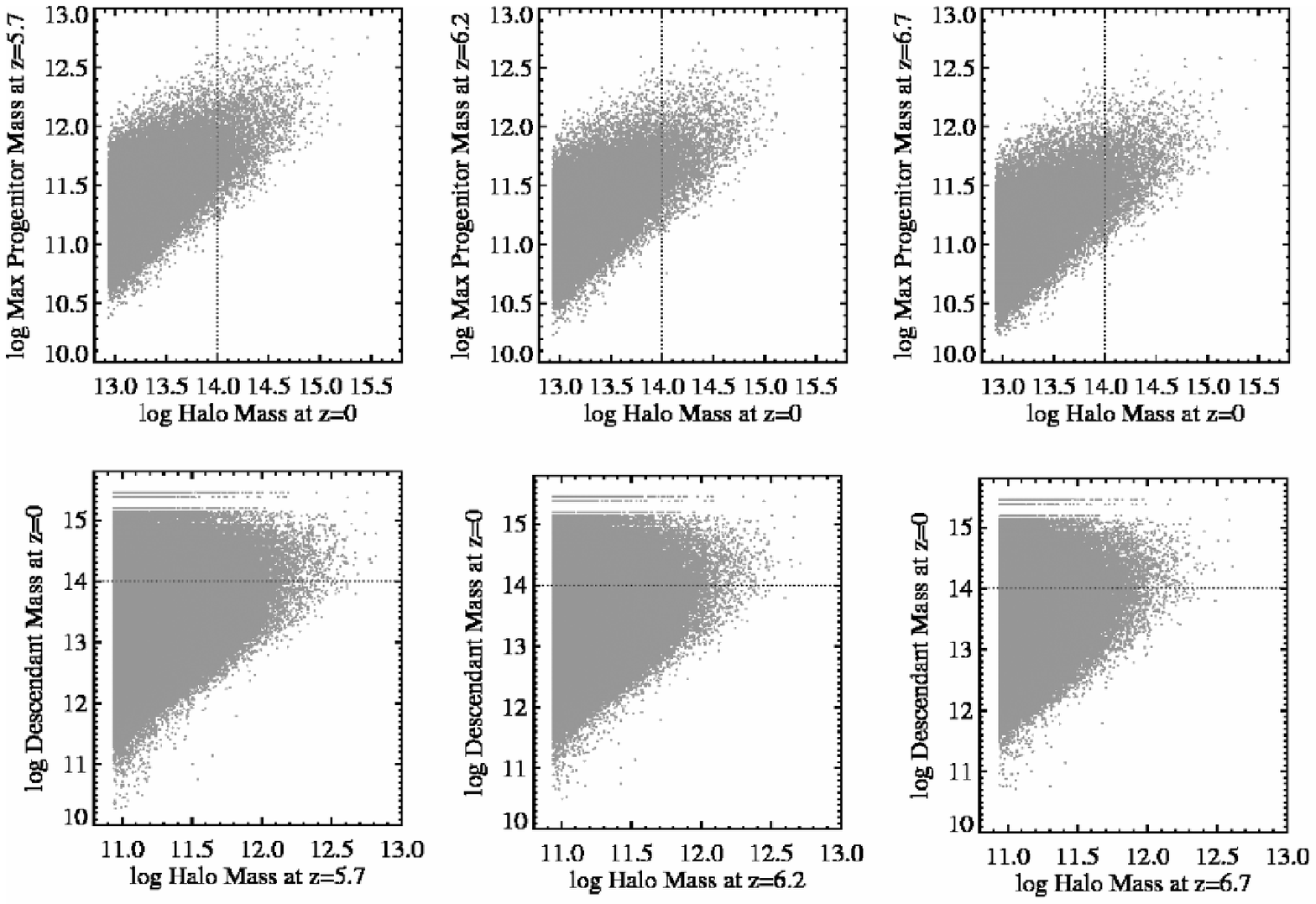}
\end{center}
\caption{\label{fig:hh}The correspondence between the most massive
  halos selected at $z=0$ and $z=6$ \citep[see also][]{trenti08}.  In
  the top row of panels we plot the mass of the most massive
  progrenitor of halos selected at $z=0$ for snapshots at $z=5.7$
  (left), $z=6.2$ (middle) and $z=6.7$ (right). In the bottom row of
  panels we plot the mass of the most massive $z=0$ descendant for
  halos selected at $z=5.7$ (left), $z=6.2$ (middle) and $z=6.7$
  (right). In all panels the dotted line indicates the mass
  corresponding to the threshold we use to define clusters at $z=0$
  ($M\ge10^{14}$ $h^{-1}$ $M_\odot$ Mpc). The dispersion in the mass
  of the most massive $z\sim6$ progenitors of $z=0$ clusters is over
  an order of magnitude. Conversely, the most massive halos present at
  $z\sim6$ are not necessarily the most massive halos at $z=0$, and a
  minority does not even pass the threshold imposed for qualifying as
  a $z=0$ cluster.}
\end{figure*}

Finally, we investigate the level of mutual correspondence between the
most massive halos selected at $z=6$ and $z=0$. In
Fig. \ref{fig:z6panels} we already saw that the richest regions are
associated with a very large number of galaxies that will become part
of a cluster when evolved to $z=0$. In the top row of
Fig. \ref{fig:hh} we show the mass of the most massive progenitors at
$z=5.7$ (left), $z=6.2$ (middle) ad $z=6.7$ (right) of halos selected
at $z=0$ \citep[see also][]{trenti08}. The dotted line indicates the
threshold corresponding to massive galaxy clusters at $z=0$. Although
the progenitor mass increases systematically with increasing local
mass, the dispersion in the mass of the most massive $z\sim6$
progenitors is about or over one order of magnitude, and this is true
even for the most massive clusters. As explained in detail by
\citet{trenti08} this observation leads to an interesting complication
when using the refinement technique often used to simulate the most
massive regions in the early Universe by resimulating at high redshift
the most massive region identified at $z=0$ in a coarse grid
simulation. In the bottom panels of Fig. \ref{fig:hh} we show the
inverse relation between the most massive halos selected at $z\sim6$
and their most massive descendant at $z=0$. From this it becomes clear
that eventhough the most massive $z\sim6$ halos (e.g. those associated
with QSOs) are most likely to end up in present-day clusters, some
evolve into only modest local structures more compatible with, e.g.,
galaxy groups. This implies that the present-day descendants of some
of the first, massive galaxies and supermassive black holes must be
sought in sub-cluster environments.

\section{Discussion}

Although our findings of the previous Section show that the apparent
lack of excess neighbour counts near $z\sim6$ QSOs is not inconsistent
with them being hosted by supermassive dark matter halos as suggested
by their low co-moving abundance and large inferred black hole mass,
it is interesting to note that none of the QSO fields have densities
that would place them amongst the richest structures in the $z\sim6$
Universe. This leads to an intriguing question: where is the
large-scale structure associated with QSOs?

One possibility that has been discussed \citep[e.g.][]{kim08} is that
while the dark matter density near the QSOs is significantly higher
compared to other fields, the strong ionizing radiation from the QSO
may prohibit the condensation of gas thereby suppressing galaxy
formation. Although it is not clear how important such feedback
processes are exactly, we have found that proto-clusters in the MR
form inside density enhancements that can extend up to many tens of
Mpc in size. Although we do not currently know whether the $z\sim6$
QSOs might be associated with overdensities on scales larger than a
few arcminutes as probed by the ACS, it is unlikely that the QSO
ionization field will suppress the formation of galaxies on such large
scales \citep{wyithe05}. An alternative, perhaps more likely,
explanation for the deficit of \ip-dropouts near QSOs, is that the
dark matter halo mass of the QSOs is being greatly
overestimated. \citet{willott05} suggest that the combination of the
steepness of the halo mass function for rare high redshift halos on
one hand, combined with the sizeable intrinsic scatter in the
correlation between black hole mass and stellar velocity dispersion or
halo mass at low redshift on the other hand, makes it much more
probable that a $10^9$ $M_\odot$ black hole is found in relatively low
mass halos than in a very rare halo of extremely high mass.
Depending on the exact value of the scatter, the typical mass of a
halo hosting a $z\sim6$ QSO may be reduced by $\sim0.5-1.5$ in log
$\cal{M}$$_{halo}$ without breaking the low redshift
$\cal{M}$--$\sigma_v$ relation. The net result is that QSOs occur in
some subset of halos found in substantially less dense environments,
which may explain the observations.  This notion seems to be confirmed
by the low dynamical mass of $\sim5\times10^{10}$ $M_\odot$ estimated
for the inner few kpc region of SDSS J1148+5251 at $z=6.43$ based on
the CO line emission \citep{walter04}. This is in complete
contradiction to the $\sim10^{12}$ $M_\odot$ stellar mass bulges and
$\sim10^{13}$ $M_\odot$ mass halos derived based on other arguments. If
true, models should then explain why the number density of such QSOs
is as observed.  On the other hand, recent theoretical work by
\citet{dijkstra08} suggests that in order to facilitate the formation
of a supermassive ($\sim10^9$ $M_\odot$) by $z\sim6$ in the first
place, it may be required to have a rare pair of dark matter halos
($\sim10^{13}$ $M_\odot$) in which the intense UV radiation from one
halo prevents fragmentation of the other so that the gas collapses
directly into a supermassive blackhole. This would constrain the QSOs
to lie in even richer environments.

\section{Recommendations for future observations}

The predicted large-scale distributions of \ip-dropouts and
\lya\ emitters as shown in, e.g., Figs. \ref{fig:field}, \ref{fig:ota} and
\ref{fig:ouchi} show evidence for variations in the large-scale
structure on scales of up to $\sim$1-2\degr, far larger than currently
probed by deep HST or large-area ground-based surveys. A full
appreciation of such structures could be important for a range of
topics, including studies of the luminosity function at $z\sim6$ and 
studies of the comparison between $\Lambda$CDM predictions and
gravitational clustering on very large scales. The total area probed by our
simulation is a good match to a survey of $\simeq20$ degree$^2$
targeting \ip-dropouts and \lya\ emitters at $z\sim6$ planned with the
forthcoming Subaru/HyperSurpimeCam (first light expected 2013;
M. Ouchi, private communications, 2008).  

We found that the \ip-dropouts associated with proto-clusters are
almost exclusively found in regions with positive density
enhancements. A proper understanding of such dense regions may also be
very important for studies of the epoch of reionization. Simulations
suggest that even though the total number of ionizing photons is much
larger in very large proto-cluster regions covering several tens of
comoving Mpc as compared to the field \citep[e.g.][but see
\citet{iliev06}]{ciardi03}, they may still be the last to fully
re-inionize, because the recombination rates are also much higher.
If regions associated with QSOs or other structures were to contain significant
patches of neutral hydrogen, this may affect both the observed number
densities and clustering of LBGs or \lya\ emitters relative to our
assumed mean attenuation \citep{mcquinn07}. However, since our work
mostly focuses on $z\approx6$ when reionization is believed to be
largely completed, this may not be such an issue compared to surveys
that probe earlier times at $z\gtrsim7$ \citep[e.g.][]{kashikawa06,mcquinn07}.

Our evaluation of the possible structures associated with QSOs leads
to several suggestions for future observations.  While it is unlikely
that the Wide Field Camera 3 (WFC3) to be installed onboard HST in
early 2009 will provide better contraints than HST/ACS due to its
relatively small field-of-view, we have shown that either by surveying
a larger area of $\sim$10\arcmin$\times$10\arcmin\ or by going deeper
by $\sim$1 mag in \zp, one significantly reduces the shot noise in the
neighbour counts allowing more reliable overdensities and (lower)
limits on the halo masses to be estimated. A single pointing with ACS
would require $\sim15-20$ orbits in \zp\ to reach a point source
sensitivity of 5$\sigma$ for a \zp=27.5 mag object at $z\sim6$. Given
the typical structure sizes of the overdense regions shown in
Fig. \ref{fig:z6panels}, a better approach would perhaps be to expand
the area of the current QSO fields by several more ACS pointings at
their present depth of \zp=26.5 mag for about an equal amount of
time. However, this may be achieved from the ground as well using the
much more efficient wide-field detector systems. Although this has
been attempted by \citet{willott05} targeting three of the QSO fields,
we note that their achieved depth of \zp=25.5 was probably much too
shallow to find any overdensities even if they are there. We would
like to stress that it is extremely important that foreground
contamination is reduced as much as possible, for example by combining
the observations with a deep exposure in the \vp\ band. This is
currently not available for the QSO fields, making it very hard to
calculate the exact magnitude of any excess counts present.  While a
depth of \zp=27.5 mag seems out of reach for a statistical sample with
HST, narrow-band \lya\ surveys targeting the typically UV-faint \lya\
emitters from the ground would be a very efficient
alternative. Although a significant fraction of sources lacking \lya\
may be lost compared to dropout surveys, they have the clear
additional advantage of redshift information. Most \lya\ surveys are
carried out in the atmospheric transmission windows that correspond to
redshifted \lya\ either at $z\approx5.7$ or $z\approx6.6$ for which
efficient narrow-band filters exist. We therefore suggest that the
experiment is most likely to succeed around QSOs at $z\approx5.7$
rather than the QSOs at $z\simeq5.8-6.4$ looked at so far. It is,
however, possible to use combinations of, e.g., the $z\approx5.7$
narrow-band filter with medium or broad band filters at $\sim$9000\AA\
to place stronger constaints on the photometric redshifts of
\ip-dropouts in QSO fields \citep[e.g., see][]{ajiki06}.

In the next decade, {\it JWST} will allow for some intriguing further
possibilities that may provide some definite answers: Using the target
0.7--0.9$\mu$m sensitivity of the Near Infrared Camera (NIRCam) on
{\it JWST} we could reach point sources at 10$\sigma$ as faint as
\zp=28.5 mag in a 10,000 s exposure, or we could map a large
$\sim$10\arcmin$\times$10\arcmin\ region around QSOs to a depth of
\zp=27.5 mag within a few hours. The Near Infrared Spectrograph
(NIRSpec) will allow $>$100 simultaneous spectra to confirm the
redshifts of very faint line or continuum objects over a $>$9 arcmin$^2$
field of view.

\section{Summary}
\label{sec:discussion}

\noindent
The main findings of our investigation can be summarized as follows.

\noindent
$\bullet$ We have used the $N$-body plus semi-analytic modeling of
\citet{delucia07} to construct the largest (4$\degr\times4\degr$) mock
galaxy redshift survey of star-forming galaxies at $z\sim6$ to
date. We extracted large samples of \ip-dropouts and \lya\ emitters
from the simulated survey, and showed that the main observational
(colours, number densities, redshift distribution) and physical
 properties ($M_*$, SFR, age, $M_{halo}$) are in fair agreement with
the data as far as they are constrained by various surveys.

\noindent
$\bullet$ The present-day descendants of \ip-dropouts (brighter than
$M^*_{UV,z=6}$) are typically found in group environments at $z=0$
(halo masses of a few times $10^{13}$ $M_\odot$). About one third of
all \ip-dropouts end up in halos corresponding to clusters, implying
that the contribution of ``proto-cluster galaxies'' in typical
\ip-dropout surveys is significant.

\noindent
$\bullet$ The projected sky distribution shows significant variations
in the local surface density on scales of up to 1$\degr$, indicating
that the largest surveys to date do not yet probe the full range of
scales predicted by our $\Lambda$CDM models. This may be important for
studies of the luminosity function, galaxy clustering, and the epoch
of reionization.

\noindent
$\bullet$ We present counts-in-cells frequency distributions of the
number of objects expected per 3.4\arcmin$\times$3.4\arcmin\ HST/ACS
field of view, finding good agreement with the GOODS field
statistics. The largest positive deviations are due to structures
associated with the seeds of massive clusters of galaxies
(``protoclusters''). To guide the interpretation of current and future
HST/ACS observations, we give the probabilities of randomly finding
regions of a given surface overdensity depending on the presence or
absence of a protocluster.

\noindent
$\bullet$ We give detailed examples of the structure of proto-cluster
regions. Although the typical separation between protocluster galaxies
does not reach beyond $\sim$10\arcmin\ (25 Mpc comoving), they sit in
overdensities that extend up to 30\arcmin\ radius, indicating that the
proto-clusters predominantly form deep inside the largest filamentary
structures. These regions are very similar to two proto-clusters of
\ip-dropouts or \lya\ emitters found in the SDF \citep{ota08} and SXDF
\citep{ouchi05} fields.

\noindent
$\bullet$ We have made a detailed comparison between the number counts
predicted by our simulation and those measured in fields observed with
HST/ACS towards luminous $z\sim6$ QSOs from SDSS, concluding that the
observed fields are not particularly overdense in neighbour counts. We
demonstrate that this does not rule out that the QSOs are in the most
massive halos at $z\sim6$, although we can also not confirm it. 
We discuss the possible reasons and implications of this intriguing
result (see the Discussion in Section 6).

\noindent
$\bullet$ We give detailed recommendations for follow-up observations
using current and future instruments that can be used to better
constrain the halo masses of $z\sim6$ QSOs and the variations in the
large-scale structure as probed by \ip-dropouts and \lya\ emitters (see Section 7).

\section*{Acknowledgments}

  Many colleagues have contributed to this work. We thank Tom Abel,
  J\'er\'emy Blaizot, Bernadetta Ciardi, Soyoung Kim, Sangeeta Malhotra, James Rhoads,
  Massimo Stiavelli, and Bram Venemans for their time and
  suggestions. We are grateful to Masami Ouchi for a careful reading
  of our manuscript and insightful comments. We owe great gratitude to
  Volker Springel and Simon White and their colleagues at MPA
  responsible for the Millennium Run Project.  The Millennium
  Simulation databases used in this paper and the web application
  providing online access to them were constructed as part of the
  activities of the German Astrophysical Virtual Observatory. RAO
  acknowledges the support and hospitality of the Aspen Center for
  Physics where part of this research was carried out.

\begin{table*}
\begin{center}
\caption{\label{tab:surveys}Overview of $i$-dropout surveys.}
\begin{tabular}{lccl}
\hline
Field Name & \multicolumn{1}{c}{Survey Area} & \multicolumn{1}{c}{$z$-band detection limit$^a$}  & \multicolumn{1}{l}{Reference}\\
 &  \multicolumn{1}{c}{(arcmin$^{2}$)} & \multicolumn{1}{c}{(AB mag)} & \\
MR mock                   & 70,000      & $\sim$27.5                  & This paper\\
\hline
HUDF                      & 11.2        & $\sim$29.2 ($10\sigma$,$0\farcs2$) & \citet{bouwens06}\\
HUDF05                    & 20.2        & $\sim$28.9 ($5\sigma$,$0\farcs2$) & \citet{bouwens07,oesch07}\\
HUDF-Ps                   & 17.0        & $\sim$28.5 ($10\sigma$,$0\farcs2$) & \citet{bouwens06}\\
GOODS                     & 316         & $\sim$27.5 ($10\sigma$,$0\farcs2$) & \citet{bouwens06}\\
ACS/GTO                   & 46          & $\sim$27.3 ($6\sigma$,$1\farcs5$) & \citet{bouwens03}\\
SDF                       & 876         & $\sim$26.6 ($3\sigma$,$2\farcs0$) & \citet{kashikawa04}\\
SXDF                      & $\sim$4,680 & $\sim$25.9 ($5\sigma$,$2\farcs0$) & \citet{ota05}\\
UKIDSS UDS + SXDF         & $\sim$2,160 & $\sim$25.0 ($5\sigma$,$2\farcs0$) & \citet{mclure06}\\
\hline
QSO SDSSJ0836+0054 ($z=5.82$) & 11.5       & $\sim$26.5 ($5\sigma$,$0\farcs2$) & \citet{zheng06,ajiki06}\\
QSO SDSSJ1306+0356 ($z=5.99$) & $\sim$11.5 & $\sim$26.5 ($5\sigma$,$0\farcs2$) & \citet{kim08}\\
QSO SDSSJ1630+4012 ($z=6.05$) & $\sim$11.5 & $\sim$26.5 ($5\sigma$,$0\farcs2$) & \citet{kim08}\\
QSO SDSSJ1048+4637 ($z=6.23$) & $\sim$30   & $\sim$26.2 ($3\sigma$,$1\farcs5$) & \citet{willott05}\\
                              & $\sim$11.5 & $\sim$26.5 ($5\sigma$,$0\farcs2$) & \citet{kim08}\\
QSO SDSSJ1030+0524 ($z=6.28$) & $\sim$30   & $\sim$26.2 ($3\sigma$,$1\farcs5$) & \citet{willott05}\\
                              & 11.5       & $\sim$26.5 ($5\sigma$,$0\farcs2$) & \citet{stiavelli05,kim08}\\
QSO SDSSJ1148+5251 ($z=6.43$) & $\sim$30   & $\sim$26.2 ($3\sigma$,$1\farcs5$) & \citet{willott05}\\
                              & $\sim$11.5 & $\sim$26.5 ($5\sigma$,$0\farcs2$) & \citet{kim08}\\
\hline
\end{tabular}
\end{center}
$^a$ The numbers between parentheses correspond to the significance and the diameter of a circular aperture.
\end{table*}

\begin{table*}
\begin{center}
\caption{\label{tab:surfdens}$i$-dropout surface densities in the MR mock survey and observations.}
\begin{tabular}{cccccccc}
\hline
 & \multicolumn{6}{c}{Surface Density (arcmin$^{-2}$)}\\
Magnitude & MR & MR  & MR  & MR  & MR  & B07$^a$ & O08$^a$\\
 & (total area) & (876 arcmin$^{2}$) & (320 arcmin$^{2}$) & (160 arcmin$^{2}$) & (11.5 arcmin$^{2}$)& &\\
\hline
$z^\prime<27.50$  & 2.31 & $2.36\pm0.31$ & $2.31\pm0.45$ & $2.31\pm0.52$ & $2.28\pm0.98$ &$2.18\pm0.23$ &\\
$z^\prime<27.00$  & 0.64 & $0.62\pm0.11$ & $0.63\pm0.15$ & $0.64\pm0.17$ & $0.63\pm0.38$ &$0.83\pm0.09$ &\\
$z^\prime<26.50$  & 0.18 & $0.17\pm0.03$ & $0.18\pm0.04$ & $0.16\pm0.06$ & $0.18\pm0.15$ &$0.33\pm0.04$ & $\sim$0.18\\
$z^\prime<26.00$  & 0.08 & $0.08\pm0.01$ & $0.08\pm0.02$ & $0.08\pm0.03$ & $0.08\pm0.09$ &$0.10\pm0.02$ & $\sim$0.11\\
$z^\prime<25.50$  & 0.04 & $0.04\pm0.01$ & $0.05\pm0.01$ & $0.04\pm0.02$ & $0.04\pm0.06$ &$0.03\pm0.01$ & $\sim$0.04\\
$z^\prime<25.00$  & 0.03 & $0.03\pm0.01$ & $0.03\pm0.01$ & $0.03\pm0.01$ & $0.03\pm0.05$ &$0.003\pm0.003$ & $\sim$0.01\\
\hline
\end{tabular}
\end{center}
$^a$Observed surface densities from \citet{bouwens07} and \citet{ota08}.
\end{table*}

\end{document}